\documentclass[10pt,letterpaper]{amsart} 
\usepackage[foot]{amsaddr}

\usepackage{color}
\usepackage{multicol}
\usepackage{mathrsfs}
\usepackage[letterpaper, margin=1in]{geometry}
\usepackage[overload]{empheq}
\usepackage{caption}
\usepackage{subcaption}
\usepackage{xcolor}
\definecolor{RED}{named}{red}
\usepackage{cancel}
\usepackage{algorithm}
\usepackage{algorithmic}

\usepackage{listings}
\usepackage{amsmath, amsthm, amssymb, wasysym, verbatim, bbm, graphics, booktabs}
\usepackage{enumerate}
\usepackage{url}
\usepackage{mathtools}
\usepackage[colorlinks,linkcolor=blue]{hyperref}
\usepackage{romannum}
\usepackage{xfrac}
\usepackage{float}
\usepackage{amsmath,bm}
\usepackage[title]{appendix}
\usepackage{todonotes}
\usepackage{subcaption}
\usepackage{caption}
\usepackage{siunitx}

\newtheorem{lemma}{Lemma}[section] 
\newtheorem{proposition}[lemma]{Proposition}

\newtheorem*{problem}{Problem}

\newtheorem*{maintheorem*}{Main Theorem}
\theoremstyle{definition}{}

\newcommand{\heavydot}{\mathbin{\vcenter{\hbox{\scalebox{1.8}{$\cdot$}}}}}

\newcommand{\ini}{\mathrm{ini}}

\newcommand{\pvtwo}{\partial_{v_y}}
\newcommand{\vone}{v_x}
\newcommand{\vtwo}{v_y}

\newcommand{\eyexterior}{E_y^{\mathrm{ext}}}
\newcommand{\eyexteriorinitial}{ E^{\mathrm{ext}}_{y,\ini}}
\newcommand{\bzexterior}{B_z^{\mathrm{ext}}}
\newcommand{\bzexteriorinitial}{ B^{\mathrm{ext}}_{z,\ini}}
\newcommand{\realint}{\int_\mathbb{R}}

\newcommand{\Ginit}{{G}_{\ini}}
\newcommand{\Ginitv}{{G}_{\ini,v}}
\newcommand{\Geq}{{G}_{\text{eq}}}
\newcommand{\Geqv}{{G}_{\text{eq},v}}
\newcommand{\Geqvtwo}{{G}_{\text{eq},v^2}}
\newcommand{\Geqzero}{G_{\text{eq},0}}
\newcommand{\Eyplasma}{E_y^{\mathrm{plasma}}}
\newcommand{\Bzplasma}{B_z^{\mathrm{plasma}}}
\newcommand{\Eyplasmaini}{E_{y,\ini}^{\mathrm{plasma}}}
\newcommand{\Bzplasmaini}{B_{z,\ini}^{\mathrm{plasma}}}

\newtheorem{rem}[lemma]{Remark}
\newtheorem{lem}[lemma]{Lemma}

\allowdisplaybreaks

\numberwithin{equation}{section}

\title[Control of plasma instabilities]{
Control of kinetic plasma instabilities by laser fields
}
\date{\today}

\author[N. Crouseilles]{Nicolas Crouseilles }
\address[Nicolas Crouseilles]{\newline Univ Rennes, INRIA (MINGuS), CNRS, IRMAR, UMR 6625, Rennes, France.}
\author[L. EINKEMMER]{Lukas Einkemmer}
\address[Lukas Einkemmer]{\newline Universitat Innsbruck, Innsbruck, Austria.}
\author[Q. Li]{Qin Li}
\address[Qin Li]{\newline University of Wisconsin Madison, Madison, USA.}
\author[U. Shumlak]{Uri Shumlak}
\address[Uri Shumlak]{\newline University of Washington Seattle, Seattle, USA}
\author[Y. Yue]{Yukun Yue}
\address[Yukun Yue]{ \newline University of Wisconsin Madison, Madison, USA.}

\begin{document}
 \pagenumbering{arabic}

\begin{abstract}
We study the possibility of controlling kinetic plasma instabilities by using lasers to apply external electromagnetic fields. We derive the dispersion relation for the corresponding mathematical description, a reduced Vlasov--Maxwell system, by extending the well-known Penrose condition. It is observed that, under very mild assumptions, the dispersion relation decouples into two parts. The first part is identical to the classic Penrose condition for the Vlasov--Poisson system, while the second part describes the influence of the laser on the transverse dynamics (e.g.~a Weibel instability). In particular, this means that the longitudinal dynamics (e.g.~a two-stream instability) can not be stabilized in this manner as far as linear theory is concerned. We show, however, that nonlinear effects can be used to couple the two parts and achieve effective control. This is done by determining the control parameters (i.e.~the form of the external electric and magnetic fields) by solving a PDE-constrained optimization problem.
\end{abstract}

\maketitle

\section{introduction}\label{sec:intro}

Maintaining the stability of a plasma is important in applications ranging from fusion energy to beam shaping. This is a difficult problem as plasmas tend to exhibit a range of instabilities that can (very rapidly) take the system away from the desired equilibrium configuration. In addition, for kinetic instabilities (such as the two-stream, Weibel, or bump-on-tail instability) the corresponding partial differential equations are difficult to solve and, in general, high-dimensional.

Recently, there has been both theoretical and numerical interest in controlling the Vlasov equation by some external input (mostly externally applied electric and magnetic fields). In the simplest case, the one-dimensional Vlasov--Poisson system is considered with an external electric field as the control. The PDE-constrained optimization (i.e.~numerical) approach in \cite{einkemmer2024suppressing} shows that time-independent external electric fields can be obtained that e.g. accomplish a desired beam profile or delay the onset of a two-stream instability. One can also look at this problem from the viewpoint of linear theory, which is commonly employed to study the stability of equilibria (see, e.g., \cite{HanKwan2024LinearLD,swanson2020}). The results in \cite{einkemmer2024control} show that a two-stream instability can be completely suppressed. However, the obtained external electric field is both time and space dependent and varies on time scales of the plasma frequency, which is usually very fast. It is worth emphasizing that these are results of linear theory. However, since the instability is effectively suppressed, the assumption of small perturbations around an equilibrium remain valid for the entire evolution.
 
While these results are intriguing, from a practical point of view they have a number of downsides. First, it is usually not possible to inject an arbitrary spatially varying external electric field into a plasma. Second, the resulting electric field (at least for the results of linear theory) needs to vary extremely fast in time (on the order of the plasma frequency). Third, the initial condition of the perturbation needs to be precisely known (which is often difficult to do in experiments). Some work in the literature has sought to address some of these issues. We mention, in particular,
\begin{itemize}
  \item In \cite{Qin2014} the control of a two-stream instability is achieved by modulating the velocity of a beam. This still requires a time-dependent electric field that varies on time scales of the plasma frequency; however, since the field is constant in space it can, in principle, be created simply by inducing a potential difference at the ends of the plasma. We note that the paper uses linear theory in which each beam is modeled as a fluid (i.e., not a fully kinetic approach).
  \item In \cite{Einkemmer2024d} a beam-heating problem that leads to a bump-on-tail instability is studied numerically. It is shown that if the beam is modulated the instability can be significantly reduced. This approach has the advantage that the timescale of the modulation can be large compared to the plasma frequency and no knowledge of the perturbation is required.
  \item In \cite{albi2025instantaneous} an instantaneous control strategy was used to determine a magnetic field that keeps the plasma away from the walls of the problem. The advantage of using the magnetic field as a control is that it can be generated e.g. by the field coils in a fusion reactor (although in practice there are, of course, some constraints on the magnetic field). A similar problem has been studied from a theoretical point of view in \cite{knopf2020}.
\end{itemize}

In general, the question whether a kinetic plasma can be controlled by external electric and magnetic fields in realistic situations, i.e.,~in situations where we cannot simply specify an arbitrary external electric field, has not been fully addressed. In this paper, we thus consider the control of a kinetic plasma by the electromagnetic fields of laser beams. Studying the interaction of plasmas and laser light is routinely done \cite{kruer:pop2000,shi:pop2018,tikhonchuk:book2024}. 
In the following analysis, we only consider low-intensity laser beams propagating linearly through the plasma and the impact of the laser fields on the kinetic dynamics on a spatially homogeneous plasma.
We do not consider nonlinear interactions affecting the laser beam or instabilities associated with laser-plasma interactions that result from high-intensity laser beams. 
Implicit in these limitations are the requirements that the laser energy be sufficiently small and the laser frequency must be sufficiently high.

Considering a laser beam significantly reduces the form of the external electric and magnetic fields that can be used as a control. Thus, it addresses points one and two above. We will study this both in the framework of linear theory and PDE constrained optimization. In contrast to \cite{einkemmer2024control}, we find that the two-stream instability cannot be stabilized by linear theory. In fact, as we will show, the dispersion relations decouple and thus the laser fields, as far as linear theory is concerned, has no effect on the density perturbations in the plasma. Nevertheless, we will show that in the PDE constrained optimization framework we can use nonlinear effects to achieve effective control.

\subsection{Problem Setup}\label{sec:setup}
The Vlasov--Maxwell system provides the fundamental description of non-relativistic magnetized plasmas. In this work, we consider the case of a plasma that is homogeneous in the $y$ and $z$ direction and interacts with the electromagnetic fields of a laser beam that travels along the $x$-axis. Without loss of generality, we choose the coordinate axis such that the laser's magnetic field is along the $z$ axis and the laser's electric field is along the $y$ axis. Then the 3+3 dimensional Vlasov--Poisson equations reduce to one spatial dimension {($x\in [0, L]\subset\mathbb{R}$  with $L>0$)} and two velocity dimensions {($v=(v_x,v_y)\in\mathbb{R}^2$)}. The corresponding equations of motion are given by
\begin{subequations}
\label{eq:Vlasov-Maxwell_system}
\begin{empheq}[left=\empheqlbrace]{align}
&\partial_t f + v_x \partial_x f + \left(E_x + v_y B_z\right) \partial_{v_x} f + \left(E_y - v_x B_z\right) \partial_{v_y} f = 0, \label{eq:VM1.5D_no_pert} \\
&\partial_x E_x = \rho-1 \,, \label{eq:Poisson_no_pert} \\
&\partial_t E_y = -\partial_x B_z - {\left(j_y-\langle j_y\rangle\right)}, \label{eq:tranverse_electric_field_no_pert} \\
&\partial_t B_z = -\partial_x E_y, \label{eq:magnetic_field_no_pert}
\end{empheq}
\end{subequations}
Here, $f(t,x,v_x,v_y)$ is the density of plasma at time $t$ at the phase space location $(x,v_x,v_y)$, {\( E_x(t, x) \) and \( E_y(t, x) \)} denote the $x$ and $y$ components of the electric field, and {\( B_z(t, x) \)} represents the $z$ component of the magnetic field. The two macroscopic quantities in the equations are density and current density:
\begin{equation}
\rho(t,x)=\int_{\mathbb{R}^2} f(t,x,v_x,v_y)\,dv_x\,dv_y\,,\quad\text{and}\quad {j_y(t, x)} = \int_{\mathbb{R}^2} v_y\, f(t,x,v_x,v_y)\, dv_x\, dv_y\,. \label{eq:current_traverse}
\end{equation}
They respectively affect \( E_x \) through Poisson's equation~\eqref{eq:Poisson_no_pert} and the evolution of \(E_y\). {Here, \(\langle \cdot \rangle\) denotes the spatial average of a function. Namely, $\langle j_y \rangle \;=\; \frac{1}{L} \int_{0}^{L} j_y(t,x)\,\mathrm{d}x$}. The stated equations are in dimensionless form with speed in units of the speed of light and time in units of the inverse plasma frequency.

Throughout the paper we assume periodic boundary conditions on $[0,L]$. The equations are equipped with initial data:
\[
f(t=0,x,v_x,v_y)=f_\ini(x,v_x,v_y)\,,\quad E_y(t=0,x)=E_{y,\ini}(x)\,,\quad B_{z}(t=0,x)=B_\ini(x)\,,
\]
where $\rho_\ini$ is determined from~\eqref{eq:current_traverse} using $f_\ini$. It is worth noting that $E_x$ at the initial time needs to be compatible with $f_\ini$ through~\eqref{eq:Poisson_no_pert}, while $E_y$ and $B_z$ are independently solved by the wave equation, with $f$ serving as a source in the form of $j_y$ for $E_y$. The equation's mathematical property and well-posedness structure was investigated in~\cite{asano, degond,glassey}.

Since Maxwell's equations are linear, one can decompose:
\[
E_y = \eyexterior + \Eyplasma \quad \text{and} \quad B_z = \bzexterior + \Bzplasma,
\]
with $(\eyexterior,\bzexterior)$ representing the external fields generated by the laser beam and $(\Eyplasma,\Bzplasma)$ representing the self-consistent fields generated by the plasma. Mathematically, this means:
\begin{itemize}
    \item[--] $(\Eyplasma,\Bzplasma)$ solve~\eqref{eq:tranverse_electric_field_no_pert}-\eqref{eq:magnetic_field_no_pert} self-consistently with the flux term $j_y$. They are thus considered self-generated fields. We denote their associated initial data by
    \begin{equation}
        \Eyplasma(t=0,x)=\Eyplasmaini{(x)}\,,\quad \Bzplasma(t=0,x)=\Bzplasmaini{(x)}\,.
    \end{equation}
    \item[--] $(\eyexterior,\bzexterior)$ solve~\eqref{eq:tranverse_electric_field_no_pert}-\eqref{eq:magnetic_field_no_pert} without the source $j_y$. This is the component that encodes the wave information from the external fields (i.e.~ the laser). This equation can be solved analytically if the initial values for $(\eyexterior,\bzexterior)$ are given by

\begin{equation}\label{eq:E_y_B_z_exterior_para_formula}
\begin{aligned}
 {\eyexteriorinitial(x)} &= \sum_k (a_k + d_k) \cos(k  x) + \sum_k (b_k + c_k) \sin (k x), \\
 {\bzexteriorinitial(x)} &= \sum_k (a_k - d_k) \cos(k  x) + \sum_k (b_k - c_k) \sin (k  x)\,,
\end{aligned}
\end{equation}
with $k\in k_0\mathbb{Z}$ and $k_0=\frac{2\pi}{L}$. For this initial data, each mode \(k\) is characterized by four coefficients \((a_k,b_k,c_k,d_k)\).  Physically, these fields correspond to two laser beams propagating in opposite directions. With this initial data, the exact solution assuming linear propagation of the laser beam is:

\begin{equation}
    \label{eq:Ey_A_form}
    \begin{aligned}
    \eyexterior(t,x) = \sum_{k\in k_0\mathbb{Z}} \Big({a}_k \cos\left(k (x-t)\right) + {b}_k \sin\left(k (x+t)\right) + {c}_k \sin\left(k (x-t)\right) + {d}_k \cos\left(k (x+t)\right)\Big),\\
    \bzexterior(t,x) = \sum_{k\in k_0\mathbb{Z}} \Big( {a}_k \cos\left(k (x-t)\right) - {b}_k \sin\left(k (x+t)\right) +  {c}_k \sin\left(k (x-t)\right) - {d}_k \cos\left(k (x+t)\right)\Big),
\end{aligned}
\end{equation}
\end{itemize}

The goal of this paper is to investigate whether it is possible to, by appropriately adjusting the external fields \( \eyexteriorinitial \) and \( B^{\mathrm{ext}}_{z,\ini} \), prevent instability when the initial distribution function is close to {a homogeneous equilibrium $\mu(v_x, v_y)$}, i.e., \( f_\ini \approx \mu + \text{(small perturbation)} \). More precisely:

\begin{problem}\label{prob:description}
Let \( K \subset \mathbb{Z}^+ \) be a set of Fourier modes used in the parameterization of the external fields. Determine the set of Fourier coefficients 
\begin{equation}\label{eqn:abcd}
\{(a_k, b_k, c_k, d_k)\}_{k \in k_0 K},    
\end{equation}
which define the initial external fields via \eqref{eq:E_y_B_z_exterior_para_formula}, such that the resulting solution to the 1.5D Vlasov–Maxwell system \eqref{eq:Vlasov-Maxwell_system} remains close to the equilibrium configuration.
\end{problem}

To investigate stability, we first perform a linear stability analysis around the homogeneous state \( \mu \). This leads to the derivation of a generalized Penrose condition that takes effects both due to the self-consistently generated electromagnetic field and due to the external electromagnetic field (generated by the lasers) into account.
When this generalized condition is violated, the solution may exhibit exponential growth, depending on the configuration of the initial external fields \( \eyexteriorinitial \) and \( B^{\mathrm{ext}}_{z,\ini} \). Within this linearized regime, our analysis indicates that the transverse components \( (E_y, B_z) \) can be influenced through an appropriately chosen external field and we propose a strategy for designing such initial conditions to suppress a Weibel instability. In contrast, the longitudinal component \( E_x \) completely decouples from the transverse components in the linear theory, and is thus not subject to any external control in this model. In this case, control can be only achieved by taking nonlinear effects into account. We therefore then use the numerical solution of a PDE-constrained optimization problem to identify external fields that are able to effect control.

The main contribution and outline of the paper is as follows:
\begin{itemize}
    \item We perform a detailed Fourier–Laplace analysis of the linearized 1.5D Vlasov–Maxwell with external fields \eqref{eq:Vlasov-Maxwell_system} and derive a generalized Penrose‑type stability criterion. This result extends the classical Penrose condition for the Vlasov–Poisson system to encompass both longitudinal electrostatic and transverse electromagnetic field dynamics in the linear regime as well as the external fields used as a control.
    
    \item We demonstrate that the linear control strategy can only be effective when the linear instability originates in the transverse electromagnetic fields (e.g.~for a Weibel instability), but not when the instabilities are induced by perturbations in the longitudinal direction (e.g.~a two-stream instability). We illustrate the linear control in the transverse direction for a Weibel instability, both theoretically and numerically.
    
    \item In the regime when linear control is ineffective (nonlinear regime or when perturbation induced in the longitudinal direction), we apply an optimization-based control framework. We utilize this framework to study the control problem for a two-stream instability.
\end{itemize}

\subsection{Numerical Solver}\label{subsec:forward_solver}

In this paper, we conduct extensive numerical experiments that require repeatedly solving system~\eqref{eq:Vlasov-Maxwell_system}.  Our forward solver is a recently developed {(see \cite{crouseilles_yue_sav_vlasov_maxwell})}, energy‐stable semi‐Lagrangian scheme based on the scalar auxiliary variable (SAV) approach, which ensures the preservation of the charge and the conservation of a modified energy functional in semi-discrete setting.  Unless otherwise specified, all simulations employ a grid of \(64\times256\times256\) points in \((x,v_x,v_y)\) and a time step \(\Delta t = 0.02\) (for the Weibel instability) and \(\Delta t = 0.1\) (for the two-stream instability).  The only exception is the forward process in the optimization experiments in Section~\ref{sec:nonlinear_control}, where, due to memory constraints, the forward solves are performed on a reduced grid of \(16\times32\times32\) points and the landscape plots which are computed on a grid of size \(32\times128\times128\).

In the following sections, we elaborate on the details of both linear and nonlinear control strategies for different scenarios, discussing when each method is applicable. Specifically, in Section \ref{sec:linear} we provide a comprehensive Fourier–Laplace analysis that leads to a stability condition in the linear regime and introduces a linear control strategy. Section \ref{sec:linear_control} demonstrates how this linear control strategy can be applied to suppress the Weibel instability. Finally, Section \ref{sec:nonlinear_control} addresses cases where linear control fails, detailing an optimization-based computational approach that successfully mitigates the instability through a nonlinear suppression strategy.

\section{Linear Analysis}\label{sec:linear}
In this section we perform a linear stability analysis by assuming that the distribution function $f$ is close to the desired equilibrium state $\mu$. We analyze the spectrum of the linearized PDE operator to determine the stability of the system. If the spectrum contains values with positive real parts, the solution may exhibit exponential growth. The condition on $\mu$ that prevents such instability is known as the Penrose condition.

To this end, we first linearize the system and perform a Fourier–Laplace analysis in Section~\ref{sec:FL_analysis}. In Section~\ref{subsec:reduce_stability_condition}, we then derive the resulting stability conditions, both in the general setting and under a symmetrized formulation. This sequence of analysis lays the foundation for designing initial transverse fields \( (E_{y,\ini}, B_{z,\ini}) \) that achieve control.

The linearization is conducted over the following steady state:
\[
(f,E_y,B_z) ={ \to \left(\mu,0,0\right)\,,}
\]
where {$\mu=\mu(v_x, v_y)$ is an homogeneous equilibrium satisfying} $\int_{\mathbb{R}^2} \mu(v_x,v_y)\,dv_x\,dv_y = 1$. This is a steady state, meaning the solution does not evolve if the initial data is prepared in that way. The stability of this steady state depends on how the system reacts if a small perturbation is applied. Following the strategies in~\cite{einkemmer2024control, Grenier2020LandauDF, HanKwan2024LinearLD, mouhot2011Landau}, we set:
\[
f(0,x,v_x,v_y) = \mu(v_x,v_y) + f_{\mathrm{pert}}(0,x,v_x,v_y)\,,\quad (E_y,B_z)=(E_{y,\mathrm{pert}},B_{z,\mathrm{pert}})\,,
\]
where $\heavydot_{\mathrm{pert}}$ denotes a small perturbation, meaning \(\|f_{\mathrm{pert}}\|_\infty \ll \|\mu\|_\infty\), and $\|(E_{y,\mathrm{pert}},B_{z,\mathrm{pert}})\|_\infty\ll 1$ with \(\|\heavydot\|_\infty\) denoting the supremum norm.

Plugging this new form into the original equation~\eqref{eq:Vlasov-Maxwell_system} and omitting the quadratic nonlinearity, we obtain the following linearized equation (for ease of presentation we drop the  subindex $\mathrm{pert}$ from now on):
\begin{subequations}
\label{eq:Vlasov-Maxwell_system_pert_linear}
\begin{empheq}[left=\empheqlbrace]{align}
&\partial_t f + v_x \partial_x f + \left(E_x + v_y B_z\right) \partial_{v_x}\mu + \left(E_y - v_x B_z\right) \partial_{v_y} \mu = 0, \label{eq:VM1.5D_linear} \\
&\partial_x E_x = \rho, \label{eq:Poisson_linear} \\
&\partial_t E_y = -\partial_x B_z - j_y, \label{eq:tranverse_electric_field_linear} \\
&\partial_t B_z = -\partial_x E_y. \label{eq:magnetic_field_linear}
\end{empheq}
\end{subequations}

The equation is better understood through a change of variables that follows the particle trajectories:
\[
g(t,x,v_x,v_y) = f(t, x + v_x t, v_x, v_y)\,.
\]
Then
\begin{equation}
\label{eq:partial_t_g}
    \partial_t g = \partial_t f + v_x \partial_x f\,,
\end{equation}
and
\begin{equation}
    \label{eq:density}
    \rho(t,x) = 
   {\int_{\mathbb{R}^2}} f(t,x,v_x,v_y)\,dv_x\,dv_y = 
   {\int_{\mathbb{R}^2}} g(t,x - v_x t,v_x,v_y)\,dv_x\,dv_y\,.
\end{equation}
This allows us to rewrite the linearized Vlasov–Maxwell equation \eqref{eq:VM1.5D_linear} in terms of \( g \):
\begin{equation}\label{eq:g_reformulation_VM}
\begin{aligned}
    \partial_t g(t,x,v_x,v_y) &+\Bigl[E_x\bigl(t,x + v_x t\bigr) + v_y\, B_z\bigl(t,x + v_x t\bigr)\Bigr]\, \partial_{v_x}\mu(v_x,v_y)\\[1mm]
    &+\Bigl[E_y\bigl(t,x + v_x t\bigr) - v_x\, B_z\bigl(t,x + v_x t\bigr)\Bigr]\, \partial_{v_y}\mu(v_x,v_y) = 0\,.
\end{aligned}
\end{equation}
We will now use Fourier-Laplace analysis to derive the generalized Penrose condition.

\subsection{Fourier-Laplace analysis}\label{sec:FL_analysis}
Fourier--Laplace analysis is a standard mathematical technique used to derive stability conditions in plasma models. Typically, a Fourier transform is applied in the spatial and velocity variables \( (x, v) \), while a Laplace transform is used in time \( t \). Transforming the system into the Fourier–Laplace domain makes the spectral properties of the equation more transparent, facilitating the identification of instabilities. The Fourier transform and Laplace transform are defined as \footnote{All Fourier formulas shown here are stated for the whole-space case \(x\in\mathbb{R}\). In the spatially periodic setting \(x\in[0,L]\), one replaces
\[
\int_{\mathbb{R}} f(x)\,e^{-ikx}\,dx
\;\longrightarrow\;
\sum_{k\in k_0\mathbb{Z}} \hat{f}_k\,e^{-ik x},
\]
and similarly replaces the inverse transform integral in \(k\) by a sum.}
: 

\begin{itemize}
    \item[--] Fourier transform in $x$, e.g.
$$\hat{F}(t,k) = \int_{-\infty}^{\infty} F(t,x)e^{-ikx}\,dx.$$
Accordingly, the inverse Fourier transform $\mathcal{F}^{-1}$ is:
\[
\mathcal{F}^{-1}[\hat{F}(k)](x) = \frac{1}{2\pi}\int_{-\infty}^{\infty} \hat{F}(k)e^{ikx}\,dk\,.
\]
\item[--] Fourier transform in $x$ and $v_x$, pointwise in $v_y$, i.e.,
\[
\hat{F}(t,k,m;v_y)
= \int_{-\infty}^{\infty}\int_{-\infty}^{\infty}
F(t,x,v_x,v_y)\,e^{-ikx}\,e^{-imv_x}\,\mathrm{d}x\,\mathrm{d}v_x.
\]
    \item[--] Laplace transform in time:
    \begin{equation}
    \label{eq:Laplace_transform_def}
    L[F](s) = \int_0^\infty e^{-st} F(t)\,dt.
\end{equation}
\end{itemize}

These classical definitions of Fourier transforms allow us to link \( \hat\rho \) and \( \hat{g} \). From~\eqref{eq:density}, we obtain
\begin{align}
\hat{\rho}(t,k) &= \int_{-\infty}^{\infty} \int_{\mathbb{R}^2} g(t,x-v_xt,v_x,v_y)e^{-ikx}\,dv_x\,dv_y\,dx \nonumber\\[1mm]
&= \int_{\mathbb{R}^2} \left[\int_{-\infty}^{\infty} g(t,x-v_xt,v_x,v_y)e^{-ik(x-v_xt)}\,dx\right] e^{-ikt\,v_x}\,dv_x\,dv_y \nonumber\\[1mm]
&= \int_{-\infty}^{\infty} \hat{g}(t,k,kt,v_y)\,dv_y\,. \label{eq:rho_Fourier_detailed}
\end{align}

Another key observation is that the four quantities \( (\rho, E_x, E_y, B_z) \) in~\eqref{eq:Vlasov-Maxwell_system_pert_linear} can be reduced to just two: \( (\rho, B_z) \). In particular, once \( (\rho, B_z) \) are determined, the electric field components \( (E_x, E_y) \) can be explicitly recovered. This reduction is formalized in the following lemma:

{\begin{lemma}\label{lem:rho_E_x_vs_E_y_B_z}
Given the charge density $\rho(t,x)$ and the magnetic field $B_z(t,x)$, the electric field components $E_x(t,x)$ and $E_y(t,x)$ can be explicitly expressed as:
\begin{equation}\label{eq:Ex_closed_form}
    E_x(t,x) = \mathcal{F}^{-1}\left[\frac{\hat{\rho}(t,k) (1-\delta_{k0})}{ik}\right](x)
\end{equation}
and
\begin{equation}\label{eq:Ey_closed_form}
    E_y(t,x) = \mathcal{F}^{-1}\left[\frac{i\partial_t \hat{B}_z(t,k)  (1-\delta_{k0})}{k}\right](x)\,.
\end{equation}
Here, $\delta_{k0}$ denotes the Kronecker delta, i.e., $\delta_{k0}=1$ if $k=0$ and $0$ otherwise.
\end{lemma}

\begin{proof}
Taking Fourier transforms of \eqref{eq:Poisson_linear} and \eqref{eq:magnetic_field_linear}, we directly solve for $\hat{E}_x(t,k)$ and $\hat{E}_y(t,k)$ in terms of known quantities for $k \neq 0$:
\[
\hat{E}_x(t,k) = \frac{\hat{\rho}(t,k)}{ik}, \quad \text{and} \quad \hat{E}_y(t,k) = \frac{i}{k}\partial_t \hat{B}_z(t,k).
\]
For the $k=0$ mode, we set:
\[
\hat{E}_x(t,0) = 0, \quad \hat{E}_y(t,0) = 0.
\]
Applying the inverse Fourier transform $\mathcal{F}^{-1}$ with respect to the spatial variable $k$, we obtain the closed-form solutions \eqref{eq:Ex_closed_form} and \eqref{eq:Ey_closed_form}.
\end{proof}}

With these preparations in place, we now apply the Fourier–Laplace transform to~\eqref{eq:g_reformulation_VM} to derive the dispersion relation for the pair $(\rho, B_z)$. The result is summarized in the following lemma:

\begin{lemma}\label{lem:density_current_Laplace}
For each wavenumber \(k\) and Laplace variable \(s\), the Laplace transforms of the Fourier modes of the density \(\hat\rho\) and magnetic field \(\hat B_z\) satisfy two coupled identities:
\begin{equation}
    \label{eq:rho_B_matrix_form}
    A(k,s)
    \begin{pmatrix}
        L[\hat{\rho}(\cdot,k)](s)\\[1mm]
        L[\hat{B}_z(\cdot,k)](s)
    \end{pmatrix}
    = b(k,s),
\end{equation}
where, for each \((k,s)\), the \(2 \times 2\) matrix \(A(k,s)\) is defined as
\begin{equation}\label{eqn:A_def}
    A(k,s)=\begin{pmatrix}
        1 + L[\Geq(\cdot,k)](s) & ik\,L[\Geqv(\cdot,k)](s)\\[1mm]
        ik\,L[\Geqv(\cdot,k)](s) & k^2 + s^2 + 1 - k^2\,L[\Geqvtwo(\cdot,k)](s)
    \end{pmatrix},
\end{equation}
and the vector \(b(k,s)\) is given by
\begin{equation}\label{eqn:b_def}
    b(k,s)=\begin{pmatrix}
        b_1(k,s)\\
        b_2(k,s)
    \end{pmatrix}=\begin{pmatrix}
        L[\Ginit(\cdot,k)](s)\\[1mm]
        -ik\,\hat{E}_y(0,k) + \hat{B}_z(0,k) \left[s + L[\Geqzero(\cdot,k)](s)\right] + ik\,L[\Ginitv(\cdot,k)](s)
    \end{pmatrix}.
\end{equation}
The six involved quantities are:
\[
\begin{aligned}
\Ginit(t,k) &= \int_{\mathbb{R}}\hat g(0,k,kt,v_y)\,\mathrm{d}v_y, 
&\quad 
\Ginitv(t,k) &= \int_{\mathbb{R}}v_y\,\hat g(0,k,kt,v_y)\,\mathrm{d}v_y,\\
\Geq(t,k) &= \int_{\mathbb{R}}t\,\hat\mu(kt,v_y)\,\mathrm{d}v_y,
&\quad 
\Geqv(t,k) &= \int_{\mathbb{R}}t\,v_y\,\hat\mu(kt,v_y)\,\mathrm{d}v_y,\\
\Geqvtwo(t,k) &= \int_{\mathbb{R}}t\,v_y^2\,\hat\mu(kt,v_y)\,\mathrm{d}v_y,
&\quad 
\Geqzero(t,k) &= \int_{\mathbb{R}}\hat\mu(kt,v_y)\,\mathrm{d}v_y.
\end{aligned}
\]
\end{lemma}
The lemma is seemingly complex, but the derivation is standard. We leave the proof to Appendix \ref{appen:lemma_density_Laplace} and Appendix \ref{append:lemma_current_relation}.

We note that the identities derived in Lemma~\ref{lem:density_current_Laplace} involve several auxiliary functions. Importantly, all of these functions can be explicitly computed from the equilibrium distribution $\mu$ and the initial perturbation $g(0)=f(0)$. Furthermore, the quantities appearing in $A(k,s)$ depend solely on $\mu$ and are independent of $f(0)$.

The solution  of this equation is the density-magnetic solution pair \( (\rho,B_z) \) in the $(k,s)$ domain. Namely, we solve, for each Fourier mode in space $k$, and each Laplace frequency in time $s$, for the value of the transformed quantity $\left(L[\hat{\rho}(\cdot,k)](s),L[\hat{B_z}(\cdot,k)](s)\right)$. Using inverse Fourier and inverse Laplace transforms, one can then reconstruct the solution profile for $(\rho,B_z)$.

\begin{rem}\label{rem:dispersion}
We have several comments:
\begin{itemize}
\item[--] If the initial distribution \(f(0,x,v_x,v_y)\) is symmetric in \(v_y\), then $\Ginitv(t,k)\equiv 0$.
    \item[--] The system recovers the classical 1D1D Vlasov-Poisson system in special cases. In particular, when $\mu(v_x,v_y)$ is symmetric in the $v_y$ direction, \(\Geqv \equiv 0\), then the system becomes decoupled and the first row becomes:
    \begin{equation}\label{eq:VP_formula}
L[\hat{\rho}(\cdot, k)](s)\left[1+ L[\Geq(\cdot, k)](s)\right] = L[\Ginit(\cdot, k)](s),
\end{equation}
This result agrees with equation (2.1) in \cite{einkemmer2024control}, obtained for the Vlasov–Poisson equation, thereby recovering the classical Penrose stability condition for the Vlasov–Poisson system.
\item[--] The solvability condition depends solely on the structure of $A$, whose four entries are determined solely by $\mu$. As a result, it is the equilibrium state that governs the stability of the system. This observation is the fundamental reason why the Penrose condition depends only on $\mu$.
\end{itemize}
\end{rem}

\subsection{Penrose condition and its reduced version}\label{subsec:reduce_stability_condition}
The form of the equation~\eqref{eq:rho_B_matrix_form} suggests that the solution exists if \(A(k,s)\) is invertible for every \((k,s)\), or equivalently $\left|\det A(k,s)\right| \neq 0$ for every $k$. When this happens, the solution is:
\begin{equation}
    \label{eq:rho_B_matrix_solution}
    \begin{pmatrix}
        L[\hat{\rho}(\cdot,k)](s)\\[1mm]
        L[\hat{B}_z(\cdot,k)](s)
    \end{pmatrix}
    =  A(k,s)^{-1}\,b(k,s)\,.
\end{equation}
Otherwise either $L[\hat{\rho}(\cdot,k)](s)$ or $L[\hat{B}_z(\cdot,k)](s)$ will have a singularity.

According to the inverse Mellin formula for the Laplace transform, a stable system requires singularities on the left half of the complex $s$-plane where $\Re(s)<0$. More precisely, if there exists a constant $\gamma$ such that the Laplace transform $L[F]$ has all its singularities on $\Re (s) < \gamma$, then the solution is bounded by $F(t) \lesssim e^{\gamma t}$. In our context, to ensure that the solution does not exhibit exponential growth, all singularities must lie strictly on the left half-plane $\Re (s) < 0$. This leads to the following criterion:

\medskip
\textbf{Generalized Penrose criterion:}
If $A$ defined in~\eqref{eqn:A_def} satisfies:
\begin{equation}
    \label{eq:Penrose_condition}
    \inf_{k \in k_0\mathbb{Z},\,\Re (s) \geq 0}\left|\det A(k,s)\right| \geq \kappa_0,
\end{equation}
for some constant $\kappa_0 > 0$, then the quantities \( \rho \) and \( B_z \) in the solution to~\eqref{eq:Vlasov-Maxwell_system_pert_linear} remain stable. 

\medskip
The converse of this criterion does not necessarily hold. That is, even if $A$ becomes singular at some point $(k_0, s_0)$ with $\Re (s_0)\geq 0$, i.e.~violating the Penrose condition, a linear stable solution may still exist—provided that the source term $b$ lies in the column space of $A(k_0, s_0)$.

\medskip
The system~\eqref{eq:rho_B_matrix_form} takes on a significantly reduced form if $\mu(\cdot, v_y)$ and $f(0,\cdot,\cdot, v_y)$ are symmetric in $v_y$. As pointed out in Remark~\ref{rem:dispersion}, in this situation $\Geqv(\cdot,k)=0$ and ${\Ginitv=0}$. Then the matrix $A$ becomes diagonal, and~\eqref{eq:rho_B_matrix_form} can be rewritten as
\begin{equation}
\label{eq:rho_B_matrix_reduced_form}
\begin{pmatrix}
D_1(k,s) & 0\\[1mm]
0 & D_2(k,s)
\end{pmatrix}
\begin{pmatrix}
L[\hat\rho(\cdot,k)](s)\\[1mm]
L[\hat B_z(\cdot,k)](s)
\end{pmatrix}
= b(k,s),
\end{equation}
or equivalently:
\begin{align}
D_1(k,s)\,L[\hat\rho(\cdot,k)](s) 
&= L[\Ginit(\cdot,k)]\,, 
\label{eq:rho_Penrose_reduced_form}\\[1mm]
D_2(k,s)\,L[\hat B_z(\cdot,k)](s) 
&= -ik\,\hat E_y(0,k) 
+ \hat B_z(0,k)\bigl[s + L[\Geqzero(\cdot,k)](s)\bigr]\,,
\label{eq:magnetic_Penrose_reduced_form}
\end{align}
where the two dispersion relations are:
\begin{equation}
    \label{eq:dispersion_rho}
    D_1(k,s) \coloneqq 1 + L[\Geq(\cdot,k)](s)\,,
\end{equation}
and
\begin{equation}
    \label{eq:dispersion_Bz}
    D_2(k,s) \coloneqq k^2 + s^2 + 1 - k^2\,L[\Geqvtwo(\cdot,k)](s)\,.
\end{equation}
The diagonal structure of~\eqref{eq:rho_B_matrix_reduced_form} clearly indicates that \( \rho \) and \( B_z \) can be solved independently. An interesting point to note is that this, in particular, implies that we can have e.g.~a Weibel instability in the transverse direction, while the dynamics in the longitudinal direction is Landau damped.

 The solvability condition for \( \rho \) depends solely on the root structure of \( D_1 \), while that for \( B_z \) depends only on the roots of \( D_2 \). When no roots are present in the right half-plane, the solutions can be written explicitly as
\begin{equation}
    \label{eq:soln_rho_B_z_reduced}
    L[\hat\rho(\cdot,k)](s) 
    = \frac{b_1(k,s)}{D_1(k,s)}\,,\quad \text{and} \quad
    L[\hat B_z(\cdot,k)](s) 
    = \frac{b_2(k,s)}{D_2(k,s)}\,.
\end{equation}
When a root is present -- for instance, if $D_1(k_0, s_0) = 0$ -- the corresponding component $\rho$ cannot be solved at that frequency, resulting in a singularity at $(k_0, s_0)$. Similarly, the vanishing of $D_2$ at some point introduces a singularity in the magnetic field component $B_z$.

As in the earlier argument, in order to apply Mellin's inverse formula and avoid exponential growth in time, the generalized Penrose stability condition reduces in this setting to the following pair of independent requirements:

\medskip
\textbf{Generalized Penrose criterion in the reduced setting:}
If $\mu(\cdot,v_y)$ {and $f(0,\cdot,\cdot,v_y)$} are symmetric on $v_y$, and
\begin{equation}
\label{eq:Penrose_condition_reduced}
\inf_{k\in \mathbb{R},\,\Re (s)>0} |D_1(k,s)| \ge \kappa_0,
\quad
\inf_{k\in \mathbb{R},\,\Re (s)>0} |D_2(k,s)| \ge \kappa_0,
\end{equation}
for some constant \(\kappa_0>0\), then $\rho$ and $B_z$ do not exhibit exponential growth. As before, it is important to emphasize that the converse does not necessarily hold. That is, even if the condition~\eqref{eq:Penrose_condition_reduced} fails--i.e., if $D_1$ or $D_2$ vanishes at some point $(k_0,s_0)$, exponential growth may still be absent, provided that the singularities are removable. Mathematically, if the numerators $b_1$ and $b_2$ in {\eqref{eq:soln_rho_B_z_reduced}} vanish simultaneously with $D_1$ and $D_2$, respectively, at those points, then the expressions $\frac{b_1}{D_1}$ and $\frac{b_2}{D_2}$ remain bounded, and Mellin's inverse formula continues to apply without yielding instability. Physically, it means the initial data is prepared to avoid exciting these unstable modes.

\begin{rem}
Recall that the standard Penrose condition~\cite{penrose1960electrostatic,einkemmer2024control,nicholson1983introduction} for the 1D1D Vlasov–Poisson system \cite{Grenier2020LandauDF,einkemmer2024control} is expressed as
\begin{equation}
    \label{eq:VP_Penrose}
    \inf_{k \in k_0\mathbb{Z},\,\Re(s) \geq 0}\left|D_1(k,s)\right| \geq \kappa_0,
\end{equation}
for some positive constant \(\kappa_0\). By analogy, our condition can be viewed as a generalization of the classical Penrose condition in the setting of the Vlasov–Maxwell system.
\end{rem}

\subsection{Discussion on controllability}\label{sec:discussion_control}
As discussed above, instability may arise if the generalized Penrose condition~\eqref{eq:Penrose_condition} fails, or when $\mu$ presents symmetry, its reduced counterpart~\eqref{eq:Penrose_condition_reduced} fails. The only remaining possibility for avoiding singularities in \( \rho \) and \( B_z \) is to design the source terms \( b_1 \) and \( b_2 \) so that the expressions in~\eqref{eq:soln_rho_B_z_reduced} possess cancelable roots—thereby preventing exponential growth in the macroscopic quantities. 

A key observation from~\eqref{eq:soln_rho_B_z_reduced} is that \( \rho \) is not linearly controllable by an external field in our model, while \( B_z \) potentially is. As outlined in the main problem in Section~\ref{sec:setup}, the only tunable parameters in our model are the initial external fields \( {\eyexteriorinitial(x)} \) and \( {\bzexteriorinitial(x)} \). These parameters do not appear in \( b_1 \) (defined in~\eqref{eqn:b_def}), which implies that we have no direct mechanism to eliminate singularities associated with \( D_1 \). As a result, if \( D_1 \) has zeros whose real part is positive \( \Re (s) > 0 \), the corresponding component \( \rho \) will inevitably exhibit exponential growth.

In contrast, the term \( b_2(k, s) \), as defined in~\eqref{eqn:b_def} depends on \( {\eyexteriorinitial(x)} \) and \( {\bzexteriorinitial(x)} \). This opens the possibility of adjusting the initial conditions so that the roots of \( b_2 \) coincide with those of \( D_2 \), effectively canceling the singularity and enabling the computation of~\eqref{eq:soln_rho_B_z_reduced} for \( B_z \).

This observation motivates the first control mechanism we propose:

\begin{itemize}
    \item \textbf{Linear control:} Extend the linear stability analysis (Lemma~\ref{lem:density_current_Laplace}) to guide the design of \( b_2 \) in such a way that the roots of \( D_2 \) are canceled, thereby suppressing instability in the transverse magnetic field \( B_z \).
\end{itemize}

This linear control strategy is expected to be effective in the regime where the distribution function remains close to the desired equilibrium \( \mu \).

When nonlinear effects become significant, the linear analysis described above no longer applies, and one must return to the full nonlinear Vlasov–Maxwell system~\eqref{eq:Vlasov-Maxwell_system}. At present, analytical tools for understanding these nonlinear behaviors remain limited, and the control strategy must therefore rely on numerical approaches:

\begin{itemize}
    \item \textbf{Nonlinear control:} Return to the original nonlinear Vlasov–Maxwell system~\eqref{eq:Vlasov-Maxwell_system} and formulate a PDE-constrained optimization problem to identify stabilizing controls.
\end{itemize}

These two control strategies—the linear and the nonlinear—are developed in detail in Sections~\ref{sec:linear_control} and~\ref{sec:nonlinear_control}, respectively, along with corresponding numerical results.

\section{Linear control strategy and its application in suppressing the Weibel instability}\label{sec:linear_control}

This section focuses on control in the linear regime. As discussed in Section~\ref{sec:discussion_control}, control is only feasible for mitigating exponential growth in $B_z$, the transverse field, whereas $\rho$ is not subject to control. Accordingly, throughout this section, we assume that $\rho$ is already stable and that $D_1(k, s)$ satisfies the Penrose condition~\eqref{eq:Penrose_condition_reduced}, but $D_2(k, s)$ violates~\eqref{eq:Penrose_condition_reduced}, and $B_z$ exhibits instability.

Our goal is to design a control strategy within this linear regime to eliminate the exponential growth in $B_z$. Specifically, we aim to derive an explicit prescription for the initial external fields $\eyexteriorinitial,\bzexteriorinitial$, expressed in terms of their Fourier coefficients $\{a_k, b_k, c_k, d_k\}$ as defined in~\eqref{eq:E_y_B_z_exterior_para_formula}, in such a way that the resulting expression for $b_2(k, s)$ cancels the root structure of $D_2(k, s)$. This cancellation removes the associated singularity and stabilizes the evolution of $B_z$.

The system~\eqref{eq:rho_B_matrix_reduced_form} is solvable only if $b_2=0$ whenever $D_2=0$. Specifically, suppose \(D_2(k,s)=0\) at \((k_0,s_0)\) with \(\Re (s_0)\ge0\), to ensure the solvability, we require
\begin{equation}
\label{eq:linear_control_condition_specific}
\lim_{(k,s)\to(k_0,s_0)} \frac{b_2}{D_2}(k,s)= \alpha\, 
\end{equation}
for some finite constant \(\alpha\in\mathbb{C}\). If so, the singular contribution associated with the zero of \(D_2\) is removed.

Now we are left with a freedom in the choice of the parameter $\alpha$. A special case is to set $\alpha=0$, or more aggressively, enforcing the zero condition {$b_2(k,s)=0$} for all $(k,s)$:
\begin{equation}\label{eq:linear_control_condition}
-\,i k\,\hat{E}_y(0,k) 
+ \hat{B}_z(0,k)\bigl[s + L[\Geqzero(\cdot,k)](s)\bigr] 
+ i k\,L[\Ginitv(\cdot,k)](s)
\equiv 0\,.
\end{equation}
This completely eliminates all roots of \(D_2(k,s)\) and returns $B_z\equiv 0$, a stable field.

\subsection{Choice of Initial Transverse Electromagnetic Fields}
It is important to note that the condition~\eqref{eq:linear_control_condition} is almost impossible to satisfy. Indeed, the control parameters \( \eyexteriorinitial(x) \) and \( \bzexteriorinitial(x) \) depend only on \( x \), so after applying the Fourier-Laplace transform, the only quantities we can control are functions of \( k \). However, equation~\eqref{eq:linear_control_condition} must hold for all \( (k, s) \). Manipulating two $1D$ functions to satisfy an equation in $2D$ is typically infeasible.

However, recalling that \(L[\Ginitv]=0\) if \(f(0,x,v_x,v_y)\) is even in \(v_y\), as pointed out in Remark~\ref{rem:dispersion}, the requirement above~\eqref{eq:linear_control_condition} is reduced to:
\begin{equation}\label{eq:linear_control_condition_simplified}
-\,i k\,\hat{E}_y(0,k) + \hat{B}_z(0,k)\bigl[s + L[\Geqzero(\cdot,k)](s)\bigr] \equiv 0\,.    
\end{equation}
This requirement can be satisfied. A straightforward way is to set
\begin{equation}\label{eqn:linear_stability}
\hat{E}_y(0,k) = 0,\quad \hat{B}_z(0,k) = 0,\quad \forall\,k\in \mathbb{R}\,.
\end{equation}
In this case, according to~\eqref{eq:magnetic_Penrose_reduced_form}, we have $D_2(k,s)\,L[\hat B_z(\cdot,k)](s)\equiv0$, suggesting $B_z\equiv0$ for all time. These discussions are made rigorous in the following proposition.

\begin{proposition} 
\label{prop:linear_stability_Ey_Bz}  
Assume \(f(0,x,v_x,v_y)\) is even in $v_y$ and \(L[\Ginitv]=0\). Suppose the Laplace transforms \( {L}[\hat{B}_z(\cdot,k)](s) \) and \( {L}[\Geqvtwo(\cdot,k)](s) \) are well-defined, and the dispersion function \( D_2(s,k) \) has only finitely many zeros in \( \Re(s) > 0 \) for each fixed $k$. Then, selecting the initial external laser fields 
\begin{equation}\label{eqn:linear_stability_x}
\eyexteriorinitial =- \Eyplasmaini\,, 
\quad 
\bzexteriorinitial =- \Bzplasmaini\,,
\end{equation}
ensures that \( B_z(t,x) \equiv 0 \) for all \( (t,x) \in [0,\infty)\times \mathbb{R} \).  
\end{proposition}  

\begin{proof} 
From~\eqref{eqn:linear_stability_x}, we derive $b_2(k,s)=0$ for all $(k,s)$, and thus
\[
D_2(s,k) {L}[\hat{B}_z(\cdot,k)](s) = 0\,.
\]  
By assumption, \( D_2(s,k) \) and \( {L}[\hat{B}_z(\cdot,k)](s) \) are analytic  in \( \Re(s) > 0 \) (see proof of Lemma \ref{appen:lem:vanishing_laplace_transform} 
 in Appendix \ref{appen:sec:vanishing_laplace_transform}). Since \( D_2(s,k) \) has at most finitely many zeros in \( \Re(s) > 0 \), \( B_z(t,k) = 0 \) for all \( t \geq 0 \).
\end{proof}

This result demonstrates that linear instability in \( (E_y, B_z) \) can be suppressed by canceling initial electromagnetic perturbations through externally applied laser fields. The result is not surprising. A straightforward interpretation is that if the initial transverse electromagnetic fields in the system are counteracted, the system experiences only the density perturbation, where Landau damping can naturally attenuate small perturbations. Conversely, if the transverse electromagnetic fields are nonzero, they evolve and, in general, trigger an instability. We should note that this observation is not new and was discussed in literature in different context. In~\cite{glassey} specifically, for studying the equation's well-posedness, the authors already noted that if $f(0,x,v_x,v_y)$ is even in $v_y$ and $E_y=B_z=0$ initially, they are zero for all time and $(f,E_x)$ solves the VP equation.

\subsection{Weibel Instability}\label{subsec:Weibel}
The Weibel instability is a fundamental electromagnetic instability that arises in plasmas with an anisotropic velocity distribution \cite{fried1959mechanism,weibel1959spontaneously}. It is widely used as a computational benchmark problem  for the Vlasov-Maxwell equations (see, e.g., \cite{califano1998kinetic,crouseilles2015hamiltonian, crouseilles2024exponential})  because, unlike electrostatic instabilities, it involves the exponential growth of a magnetic field until nonlinear saturation. We verify the application of the linear control strategy discussed in Proposition~\ref{prop:linear_stability_Ey_Bz} to suppress the Weibel instability below.

We consider initial data of the following form
\begin{equation}
    \label{eq:Weibel_initial}
    f_{\ini}(x,v_x,v_y) = \mu^{\mathrm{Weibel}}(v_x, v_y) \left[1 + \sum_k\alpha_k \sin(k k_0 x)+\sum_k \beta_k \cos(k k_0 x)\right], 
\end{equation}
and
\begin{equation}
\label{eq:Weibel_E_y_initial}
\begin{cases}
      \Eyplasma(0,x) ={\Eyplasmaini(x) =} \sum_k \left({a_k} \sin \bigl(k k_0 x\bigr)+b_k\cos\bigl(k k_0 x \bigr)\right)\,,\\
      \Bzplasma(0,x) = {\Bzplasmaini(x)=}\sum_k \left({c_k} \sin \bigl(k k_0 x \bigr)+d_k\cos\bigl(k k_0 x\bigr)\right)\,,
      \end{cases}
\end{equation}
with $x\in [0, L], L>0$, $k_0 = 2\pi/L$, and the equilibrium with temperature anisotropy is given by
\begin{equation}
\label{eq:Weibel_equilibrium}
    \mu^{\mathrm{Weibel}}(v_x,v_y)=\frac{1}{\pi v^2_{th}\sqrt{T_r}} \exp\!\left(-\frac{v_x^2 + \frac{v_y^2}{T_r}}{v^2_{th}}\right)\,.
\end{equation}
Here \( v_{th} \) denotes the thermal velocity, \( T_r \) is the temperature ratio, \( \alpha_k \) and \( \beta_k \) represent the strengths of the \( k \)-th Fourier mode in the distribution perturbation, while $a_k, b_k, c_k, d_k$ represent the strengths of the \( k \)-th mode for the external fields. In our computation, we use the following parameters
\begin{equation}\label{eqn:weibel_parameter}
v_{th} = 0.3, \quad T_r = 12, \quad k_0 = 1.25,\quad 
\alpha_k = 0, \quad \beta_k =
\begin{cases}
    1 \times 10^{-3}, & k = 3, \\
    0, & \text{otherwise}.
\end{cases}
\end{equation}
The initial values for the plasma-induced transverse electric and magnetic fields are set as:
\begin{equation}\label{eqn:weibel_initial_EB}
\Eyplasmaini(x) = 0, \quad \Bzplasmaini(x) = 1\times 10^{-3} \cos(3 k_0 x)\,,
\end{equation}
which corresponds to $a_k=b_k=c_k=0$ for all $k$, $d_3=1\times 10^{-3}$ and $d_k=0$ for $k\neq 3$ in \eqref{eq:Weibel_E_y_initial}.

\begin{figure}[htb]
  \centering
  \includegraphics[width=0.5\linewidth]{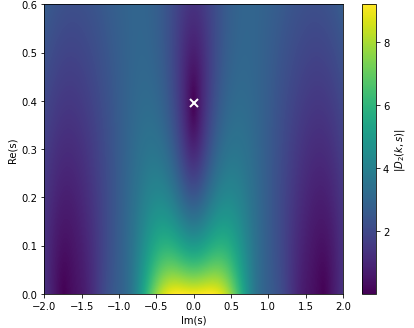}
  \caption{Magnitude of the Weibel dispersion function $D_2(k,s)$ for $k = 3$, plotted over the complex-$s$ plane.  The global minimum 
    $\lvert D_2\rvert \approx 2.643\times10^{-4}$ occurs at 
    ${\Re(s)}=0.3966$, $\text{Im}(s)=0.0$, and is marked by the white cross.}
  \label{fig:Weibel_D2_roots}
\end{figure}

This is the setting where, due to the symmetry of $\mu$, \(\Geqv\equiv0\), and $D_1(k,s)$ controls the stability of the longitudinal modes \((\rho,E_x)\) and $D_2(k,s)$ controls the stability of the transverse field \((E_y,B_z)\). Moreover, $D_1(k,s)$ satisfies the Penrose condition that ensures damping in \((\rho,E_x)\), while $D_2(k,s)$ does not, leading to an instability in \((E_y,B_z)\). {Specifically, in Figure~\ref{fig:Weibel_D2_roots}, we plot the modulus of \(\lvert D_2(k,s)\rvert\) over a grid in the complex \(s\)-plane (real and imaginary axes).  When \(k=3\), one observes a zero of \(D_2(3,s)\) with \(\Re(s)>0\) (the white cross in Figure~\ref{fig:Weibel_D2_roots}), which corresponds to a growing mode and thus indicates instability.}

\begin{figure}[ht]
    \centering
    \includegraphics[width=0.35\linewidth]{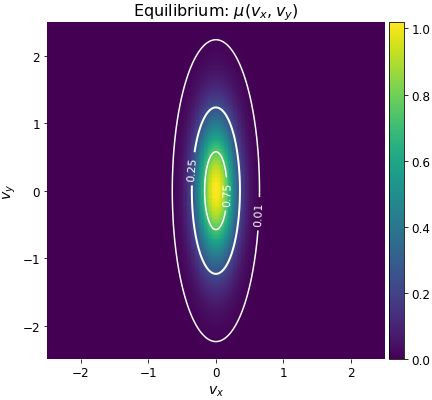}
    \includegraphics[width=0.63\linewidth]{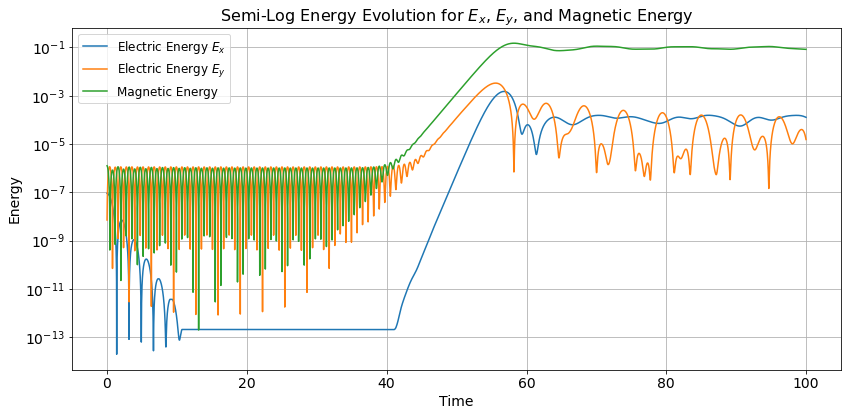}
    \caption{The panel on the left shows the equilibrium distribution using parameters in~\eqref{eqn:weibel_parameter}. The panel on the right shows, on the semi-log scale, evolution of $E_x, E_y$ and $B_z$ energy.}\label{fig:no_control_Weibel_equilibrium} 
\end{figure}

{Figure~\ref{fig:no_control_Weibel_equilibrium} illustrates the onset of the Weibel instability.  In the left panel, the equilibrium distribution \(\mu^{\mathrm{Weibel}}(v_x,v_y)\) is shown, with white isolines highlighting its anisotropic, elliptical shape. The right panel shows a semi-logarithmic plot of the time evolution of the electromagnetic energies in \(E_x\), \(E_y\), and \(B_z\) up to \(t = 100\).} Initially in the linear regime, the energy in \(E_x\), the longitudinal field, decays exponentially, while the energies in \(E_y\) and \(B_z\) grow exponentially. Once \(E_y\) and \(B_z\) reach sufficiently large amplitudes, the system enters the nonlinear regime, and \(E_x\) starts increasing as well.

\begin{figure}[ht]
  \centering
  \begin{minipage}{0.7\textwidth}
    \centering
    \includegraphics[width=\linewidth]{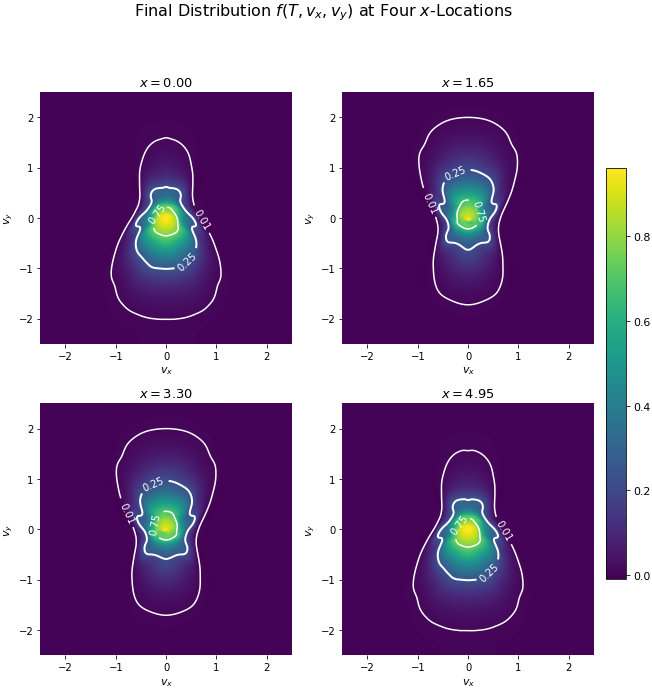}
    \vspace{0.5em}
  \end{minipage}%
  \hfill
 
  \caption{The final distribution with no control at \( t = 500 \), shown using four 2D contour slices \( f(t=500, x_i, v_x, v_y) \) at equally spaced \( x_i \).}
  \label{fig:final_distribution_no_control}
\end{figure}

\begin{figure}[htbp]
  \centering
  \begin{subfigure}[b]{0.32\linewidth}
    \centering
    \includegraphics[width=\linewidth]{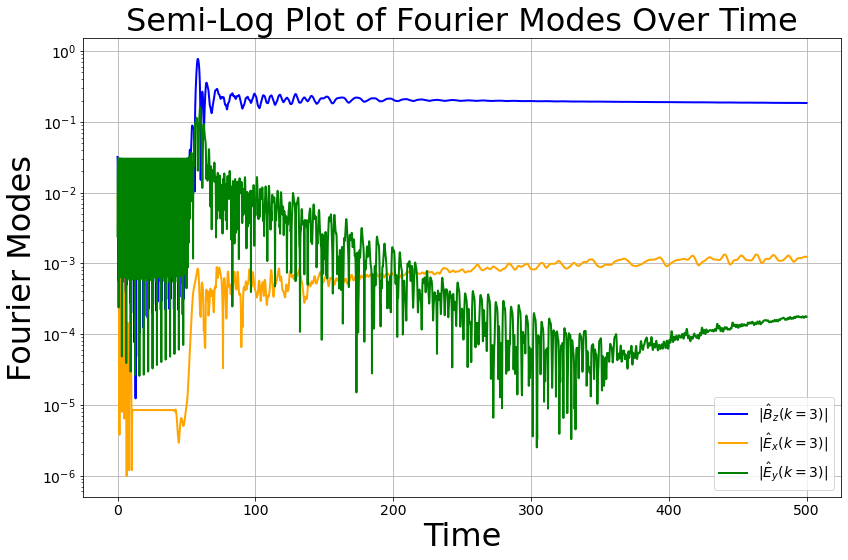}
    \caption{Uncontrolled evolution of Fourier modes $\hat E_x$, $\hat E_y$, and $\hat B_z$ for $k=3$.}
    \label{fig:no_control_electromagnetic_field_modes}
  \end{subfigure}\hfill
  \begin{subfigure}[b]{0.32\linewidth}
    \centering
    \includegraphics[width=\linewidth]{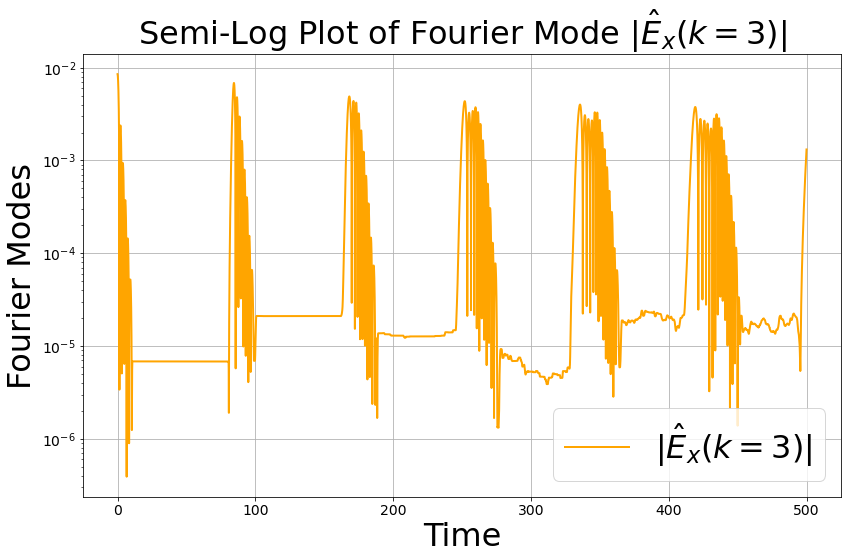}
    \caption{Fourier mode $\hat E_x$ for $k=3$ under laser stabilization.}
    \label{fig:optimal_control_electromagnetic_field_energy_evolution}
  \end{subfigure}\hfill
  \begin{subfigure}[b]{0.32\linewidth}
    \centering
    \includegraphics[width=\linewidth]{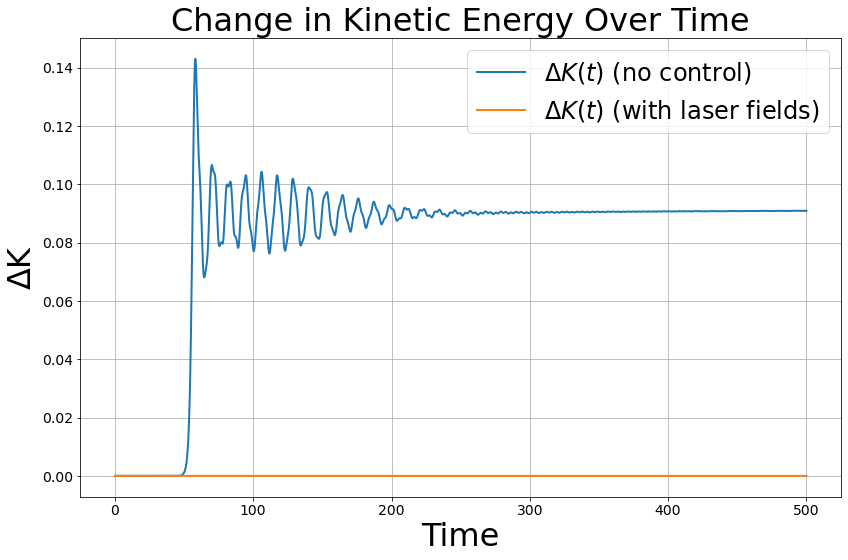}
    \caption{Comparison of the kinetic‐energy change $\Delta K(t)$.}
    \label{fig:comparison_deltaK}
  \end{subfigure}
  \caption{Comparison of uncontrolled vs.\ laser‐stabilized dynamics.}
  \label{fig:fourier_mode_comparison}
\end{figure}

To more clearly illustrate the instability, we extend the simulation up to \( t = 500 \) to capture its full development. Figure~\ref{fig:final_distribution_no_control} shows the final distribution function \( f(500, x, v_x, v_y) \) at {four} representative spatial locations. The distribution has clearly deviated from the initial equilibrium. Additionally, in Figure~\ref{fig:no_control_electromagnetic_field_modes}, we plot the time evolution of the \( k = 3 \) Fourier mode of the electromagnetic fields—specifically, \( \hat{E}_x(t, 3) \), \( \hat{E}_y(t, 3) \), and \( \hat{B}_z(t, 3) \). The results clearly indicate instability in the system, particularly in the magnetic component \( B_z \).

We now apply the linear control strategy developed above to suppress the onset of instability. Specifically, we impose external fields that satisfy the stability condition~\eqref{eqn:linear_stability_x}, with the goal of eliminating the exponential growth in \( B_z \). Based on the initial configuration in~\eqref{eqn:weibel_initial_EB}, we set: 
\[
\eyexteriorinitial(x) = 0, \qquad
\bzexteriorinitial(x) = 10^{-3} \cos(3 k_0 x).
\]
In the parameterization given by~\eqref{eq:E_y_B_z_exterior_para_formula}, this corresponds to choosing \( a_3 = 5 \times 10^{-4} \), \( d_3 = -5 \times 10^{-4} \), and setting all other Fourier coefficients \( a_k, b_k, c_k, d_k \) to zero (cf.~\eqref{eqn:abcd}).

\begin{figure}[ht]
  \centering
  \begin{minipage}{0.7\textwidth}
    \centering
    \includegraphics[width=\linewidth]{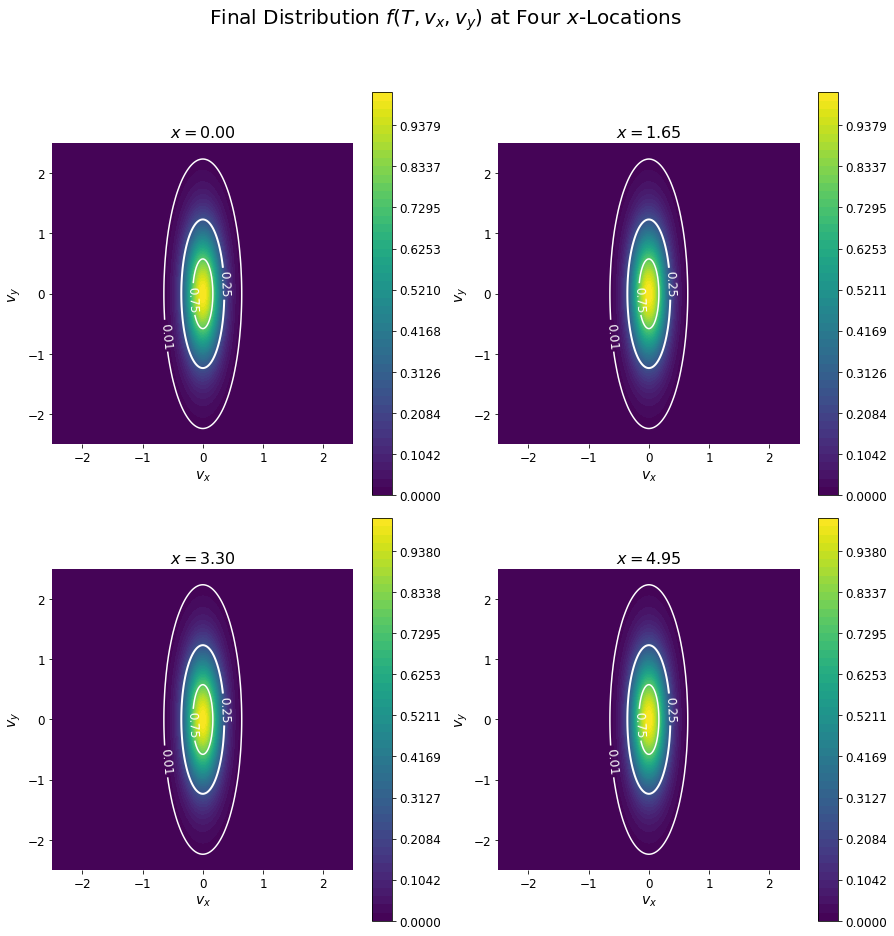}
    \vspace{0.5em}

  \end{minipage}%
  \hfill

  \caption{The final distribution with control at \( t = 500 \), shown using four 2D contour slices \( f(t = 500, x_i, v_x, v_y) \) at equally spaced \( x_i \).}
  \label{fig:final_distribution_optimal_control}
\end{figure}

We then solve the {1.5D} Vlasov–Maxwell system~\eqref{eq:Vlasov-Maxwell_system} with these newly designed external fields. Stabilization is immediate.
In Figure~\ref{fig:final_distribution_optimal_control}, we show the controlled distribution function at \( t = 500 \), which remains close to the equilibrium, in stark contrast to the uncontrolled case shown in Figure~\ref{fig:final_distribution_no_control}.
In Figure~\ref{fig:optimal_control_electromagnetic_field_energy_evolution}, we plot the time evolution of the third Fourier mode. The electric field mode \( \hat{E}_x(k=3) \) now decays, while both \( \hat{E}_y(k=3) \) and \( \hat{B}_z(k=3) \) remain identically zero throughout the simulation, confirming that our selected external fields effectively suppress the Weibel instability.

In addition, as a measure of instability, or deviation from equilibrium, we also track the change in kinetic energy, defined as the difference between the initial and current kinetic energies of the system:
\begin{equation}\label{eqn:objective}
\Delta K(t)
\;=\;-\frac{1}{2}\int_{0}^{L}\int_{\mathbb{R}^2}\bigl(v_x^2+v_y^2\bigr)\,\bigl[f(t,x,v_x,v_y)-f(0,x,v_x,v_y)\bigr]\,\mathrm{d}v_x\,\mathrm{d}v_y\,\mathrm{d}x.
\end{equation}
Figure \ref{fig:comparison_deltaK} compares the change in kinetic energy over time for the cases with and without external laser fields. It clearly shows that, with the laser fields applied, the kinetic energy remains essentially constant, whereas in the absence of external fields it undergoes a substantial change. It is worth noting that with the applied laser field, the system effectively reduces to the Vlasov–Poisson model, in which classical Landau damping occurs, providing an alternative explanation for this result.

\begin{figure}[htb]
    \centering
    \begin{minipage}{0.3\textwidth}
        \centering
        \includegraphics[width=\textwidth]{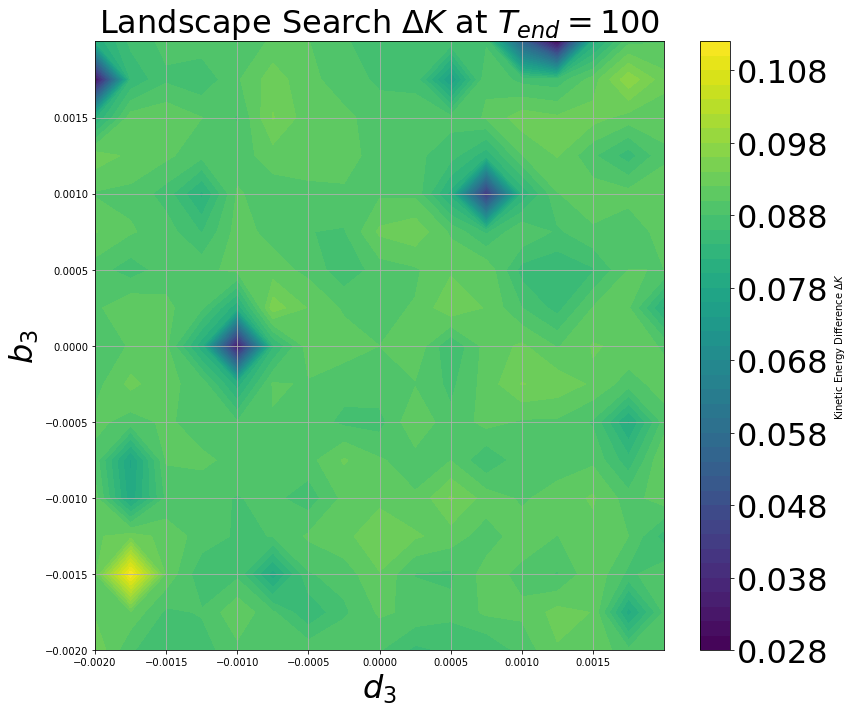}
        \caption*{(a){ $\Delta K(t=100)$}.}
    \end{minipage}
    \begin{minipage}{0.3\textwidth}
        \centering
        \includegraphics[width=\textwidth]{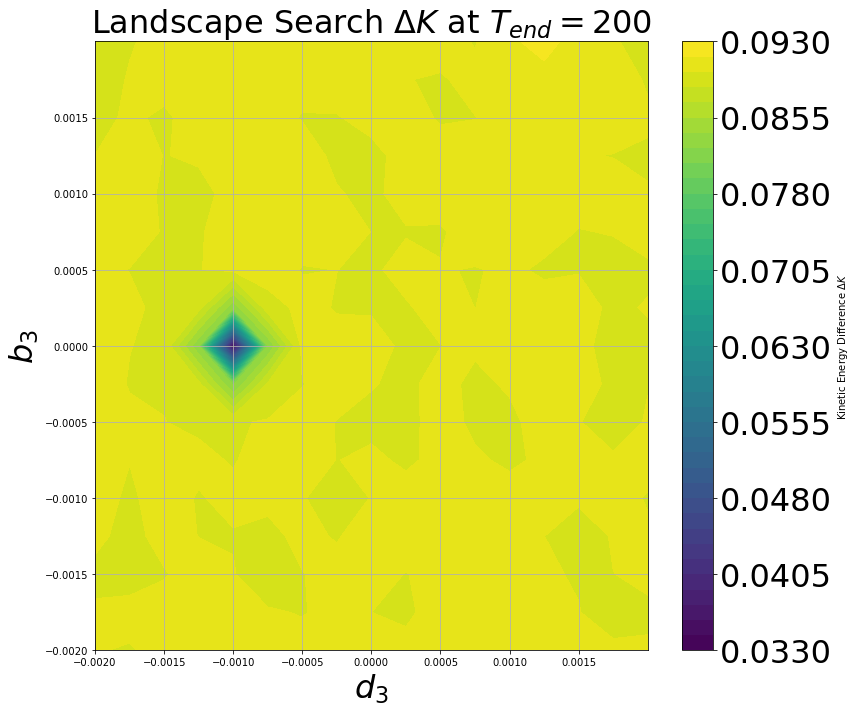}
        \caption*{(b) $\Delta K(t=200)$.}
    \end{minipage} 
    \begin{minipage}{0.3\textwidth}
        \centering
        \includegraphics[width=\textwidth]{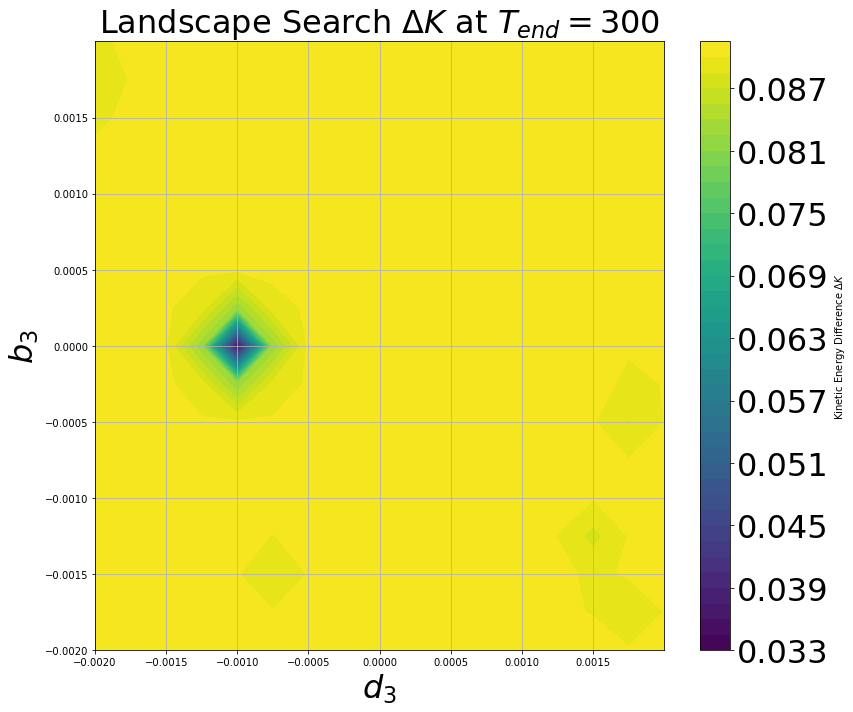}
        \caption*{(c) $\Delta K(t=300)$.}
    \end{minipage}
    
    \caption{Difference in kinetic energy {($\Delta K(t)$ (given by \eqref{eqn:objective})} 
    at various time instances for \( (d_3,b_3) \in  [-2 \times 10^{-3},\, 2 \times 10^{-3}]^2 \). }
    \label{fig:weibel_instability_landscape_search}
\end{figure}

Next, we conduct a landscape analysis over a range of external field parameters. Specifically, we consider the following form for the initial external fields: 
\[
{\eyexteriorinitial(x)} = b_3 \cos(3 k_0 x), \qquad {\bzexteriorinitial(x) }= d_3 \cos(3 k_0 x), 
\]
and examine how the system's behavior depends on the parameters \( d_3,b_3 \). We evaluate the objective function~\eqref{eqn:objective} over the parameter domain \( (d_3, b_3) \in [-2 \times 10^{-3},\, 2 \times 10^{-3}]^2 \), a range comparable to the initial amplitudes of \( (E_y^{\mathrm{ext}}, B_z^{\mathrm{ext}}) \). The numerical results are presented in Figure~\ref{fig:weibel_instability_landscape_search} at times \( t = 100 \), \( 200 \), and \( 500 \). The main observation is that only a small region around the analytically determined optimal values for $d_3$ and $b_3$ is effective in suppressing the instability.

\begin{figure}[htbp]
  \centering
  \begin{subfigure}[b]{0.4\linewidth}
    \centering
    \includegraphics[width=\linewidth]{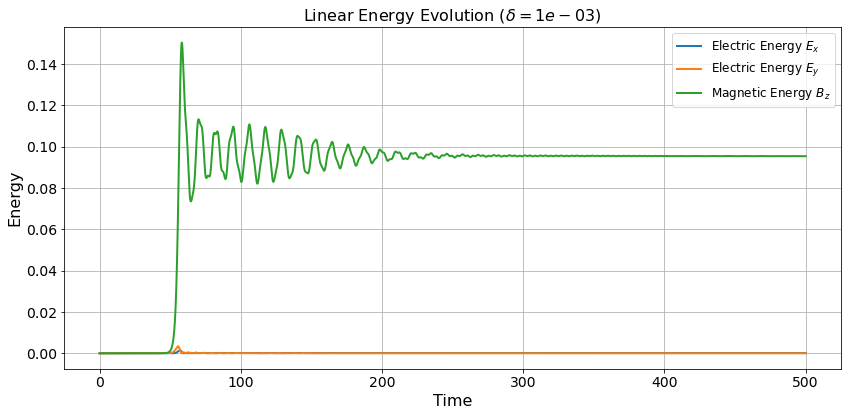}
    {(a) $\delta = 10^{-3}$.}
    \label{fig:energy_a}
  \end{subfigure}\hspace{1cm}
  \begin{subfigure}[b]{0.4\linewidth}
    \centering
    \includegraphics[width=\linewidth]{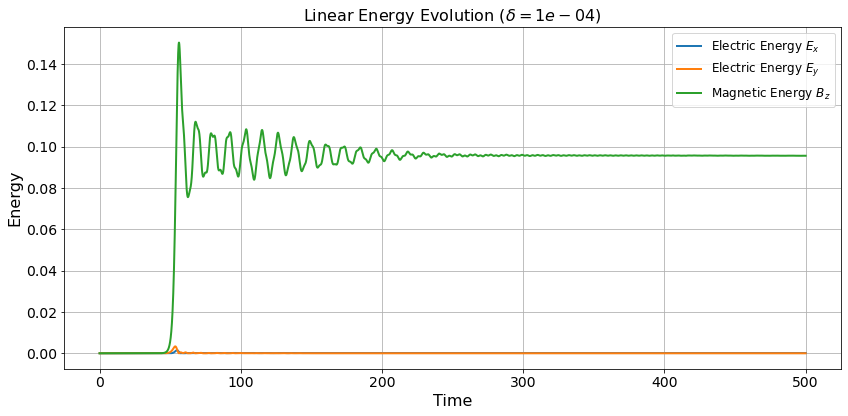}
   {(b) $\delta = 10^{-4}$.} 
    \label{fig:energy_b}
  \end{subfigure}

  \vspace{1ex}
  
  \begin{subfigure}[b]{0.4\linewidth}
    \centering
    \includegraphics[width=\linewidth]{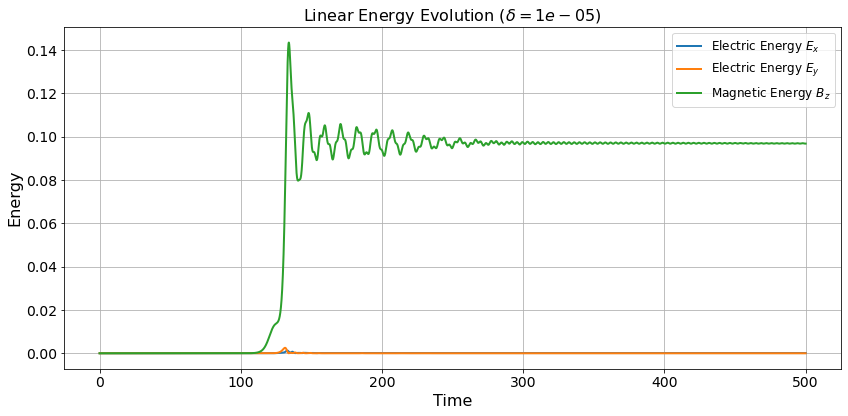}
{(c) $\delta = 10^{-5}$.}
    \label{fig:energy_c}
  \end{subfigure}\hspace{1cm}
  \begin{subfigure}[b]{0.4\linewidth}
    \centering
    \includegraphics[width=\linewidth]{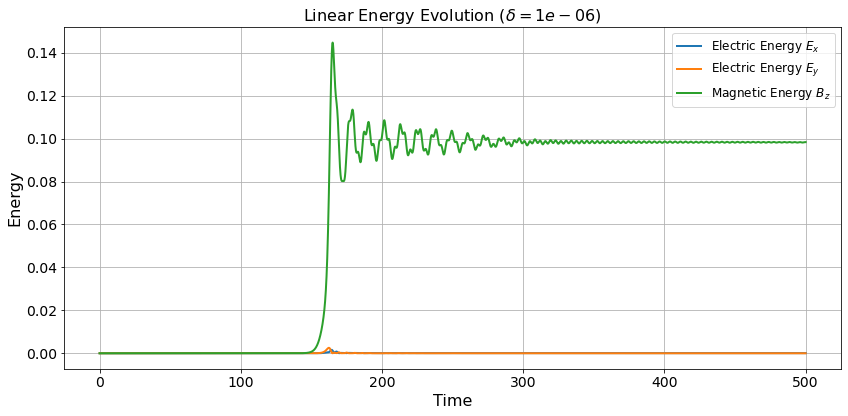}
{(d) $\delta = 10^{-6}$.} 
    \label{fig:energy_d}
  \end{subfigure}

  \begin{subfigure}[b]{0.4\linewidth}
    \centering
    \includegraphics[width=\linewidth]{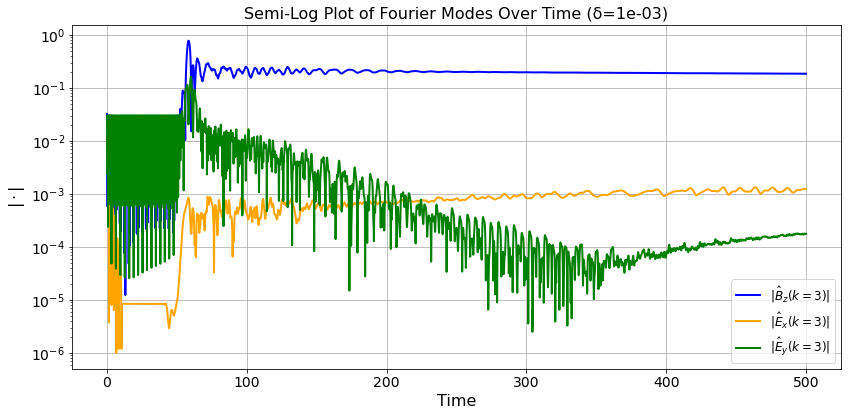}
{(e) $\delta = 10^{-3}$.}
    \label{fig:fourier_a}
  \end{subfigure}\hspace{1cm}
  \begin{subfigure}[b]{0.4\linewidth}
    \centering
    \includegraphics[width=\linewidth]{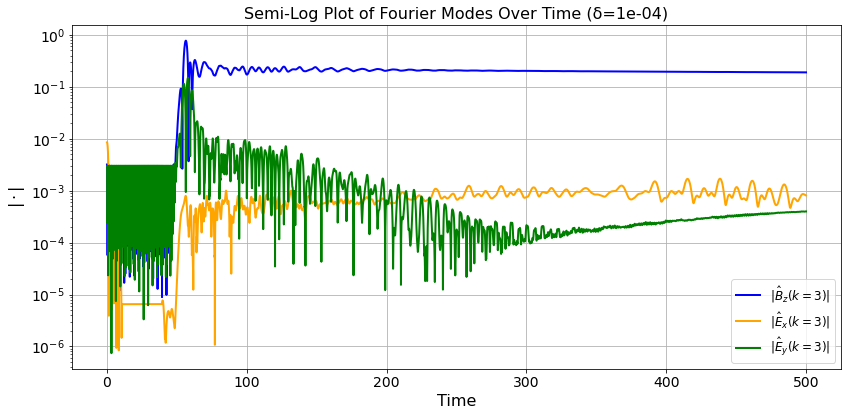}
{(f) $\delta = 10^{-4}$.}
    \label{fig:fourier_b}
  \end{subfigure}
  
  \begin{subfigure}[b]{0.4\linewidth}
    \centering
    \includegraphics[width=\linewidth]{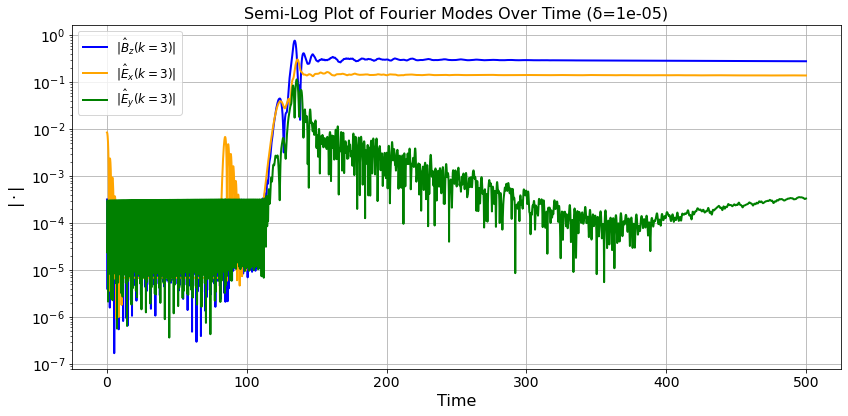}
{(g) $\delta = 10^{-5}$.}
    \label{fig:fourier_c}
  \end{subfigure}\hspace{1cm}
  \begin{subfigure}[b]{0.4\linewidth}
    \centering
    \includegraphics[width=\linewidth]{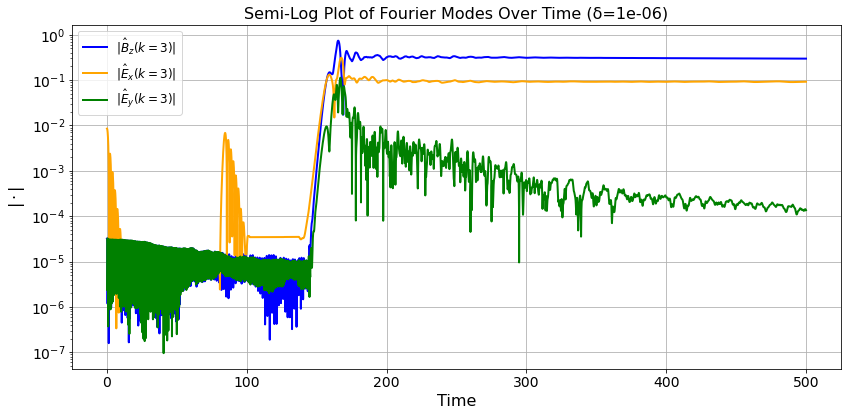}
{(h) $\delta = 10^{-6}$.}
    \label{fig:fourier_d}
  \end{subfigure}

  \caption{{(a–d)} Evolution of the electromagnetic energies for perturbation amplitudes \(\delta=10^{-3},\,10^{-4},\,10^{-5},\,10^{-6}\).  
{(e–h)} Evolution (semi‐log plots) of the Fourier modes \(\lvert\hat E_x(k=3)\rvert\), \(\lvert\hat E_y(k=3)\rvert\), and \(\lvert\hat B_z(k=3)\rvert\) under the same perturbation amplitudes.
  }
  \label{fig:energy_fourier_matrix}
\end{figure}

To investigate this in more detail we consider (small) perturbations of size $\delta$ to the optimal value predicted by linear theory. Specifically, we set
\[
E_{y, \mathrm{ini}}^{\mathrm{ext}}(x) = 0, 
\qquad
B_{z, \mathrm{ini}}^{\mathrm{ext}}(x) = \bigl(10^{-3} + \delta\bigr) \cos(3 k_0 x)
\]
with \(\delta = 10^{-3},\,10^{-4},\,10^{-5}\, \text{and }10^{-6}\). Figure~\ref{fig:energy_fourier_matrix} presents the resulting time evolution of the electromagnetic energies and their \(k=3\) Fourier modes. We can see that in contrast to Figure~\ref{fig:optimal_control_electromagnetic_field_energy_evolution}, where \(\hat E_y\) and \(\hat B_z\) remain identically zero, even small perturbations eventually induce rapid growth of the magnetic energy. The onset of the linear growth, however, is delayed and the magnitude of that delay depends on $\delta$. The time at which linear growth starts ranges from approximately $t=50$ (no control) to $t=150$ ($\delta = 10^{-6}$).

\section{Nonlinear control case and application to the two-stream instability}\label{sec:nonlinear_control}
While the linear control strategy offers valuable insights, it also has significant limitations. In particular, within linear theory we are incapable of addressing instabilities in the longitudinal direction, i.e.~affecting \( \rho \) and \( E_x \). Thus, in this section we will turn to a PDE-constrained optimization framework to obtain laser configurations that suppress instabilities in the longitudinal direction. As an illustrative example, we consider the two-stream instability.

\begin{figure}
  \centering
  \begin{minipage}{0.48\linewidth}
    \centering
    \includegraphics[width=\linewidth]{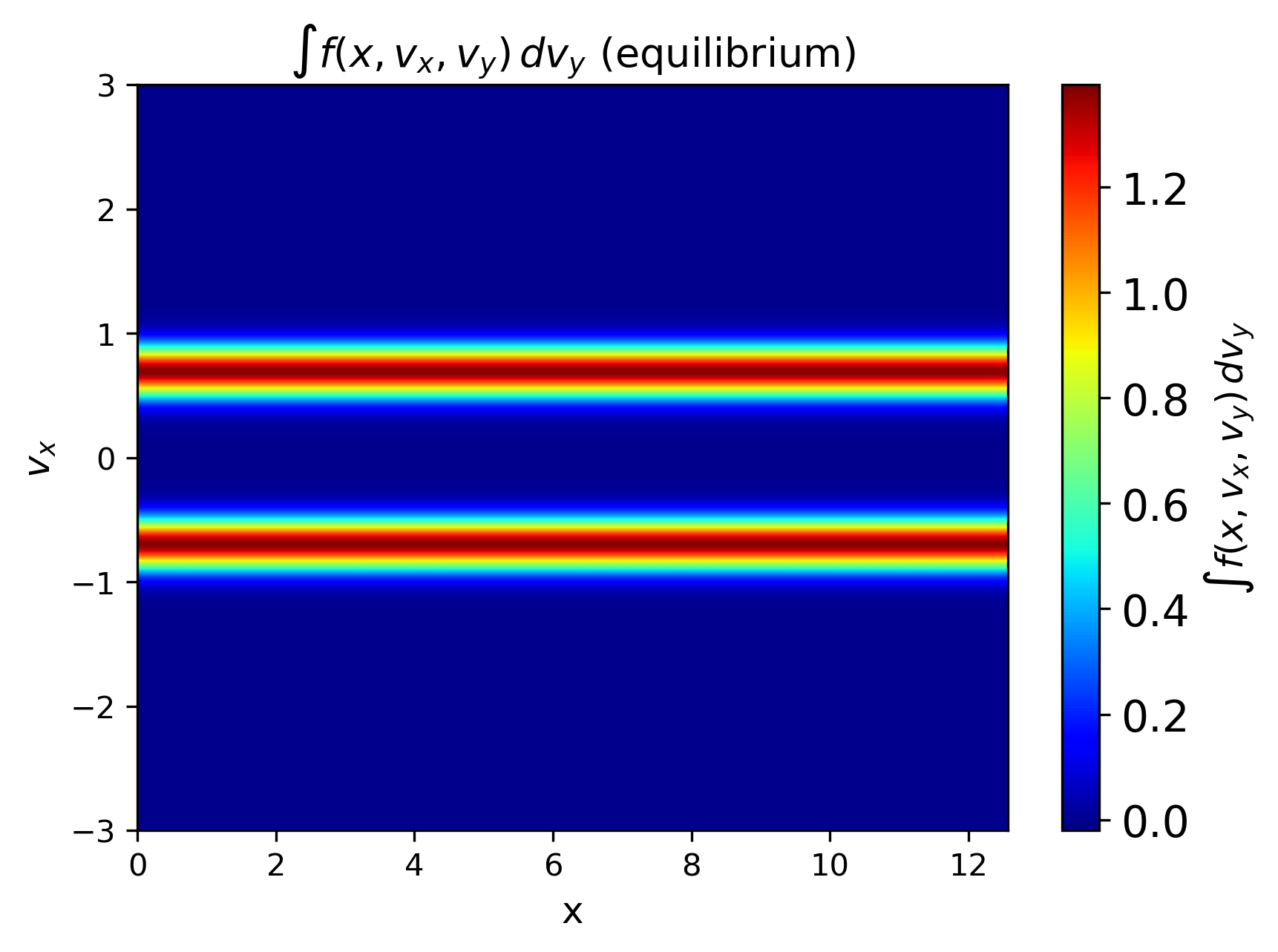}
    \vspace{0.5em}
    {(a) Equilibrium distribution.}
  \end{minipage}
  \hfill
  \begin{minipage}{0.48\linewidth}
    \centering
    \includegraphics[width=\linewidth]{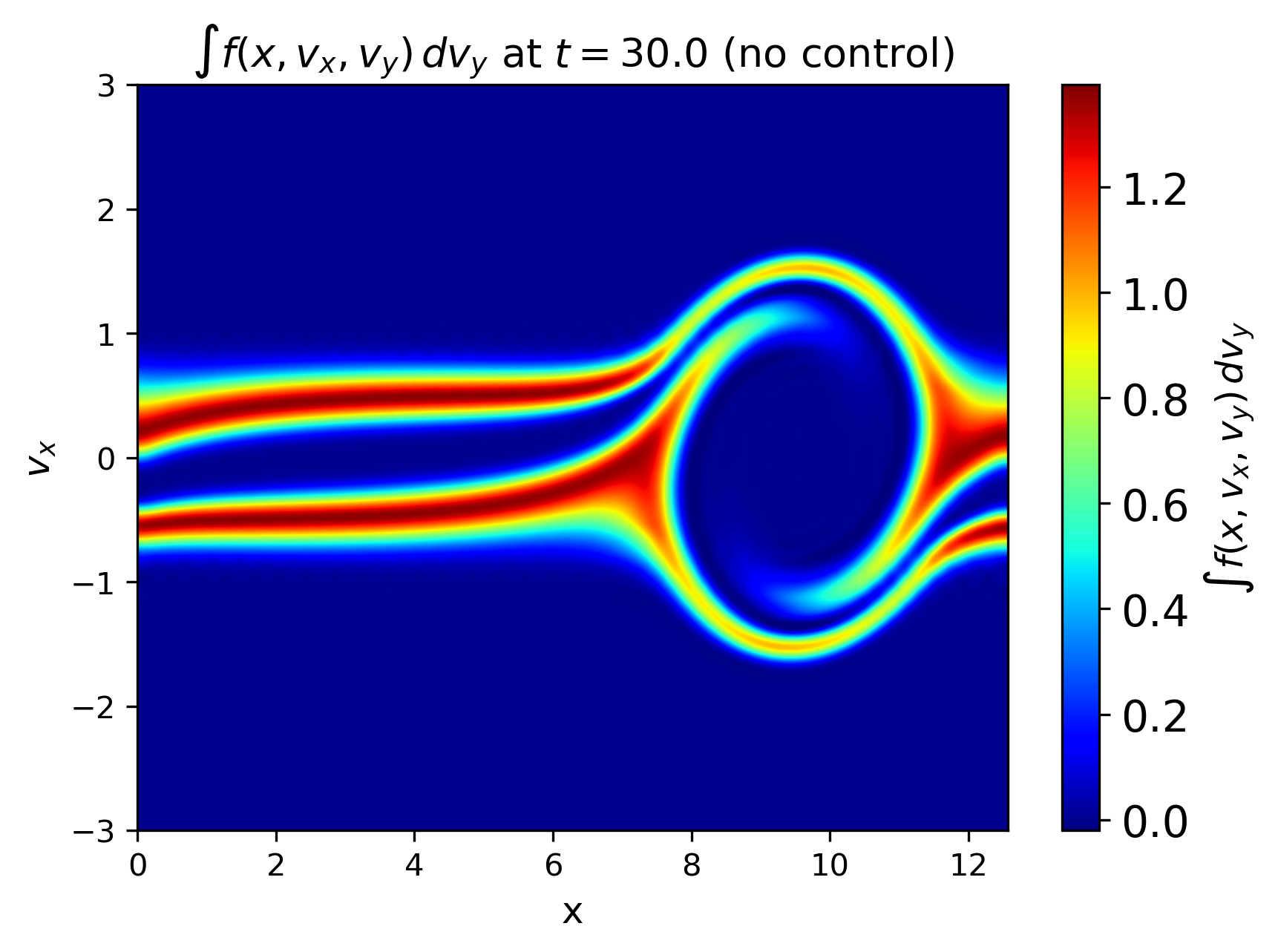}
    \vspace{0.5em}
    {(b) Distribution at \(t=30\).}
    \label{fig:t40_distribution_no_control}
  \end{minipage}
  
  \vspace{1em} 
  
  \begin{minipage}{0.48\linewidth}
    \centering
    \includegraphics[width=\linewidth]{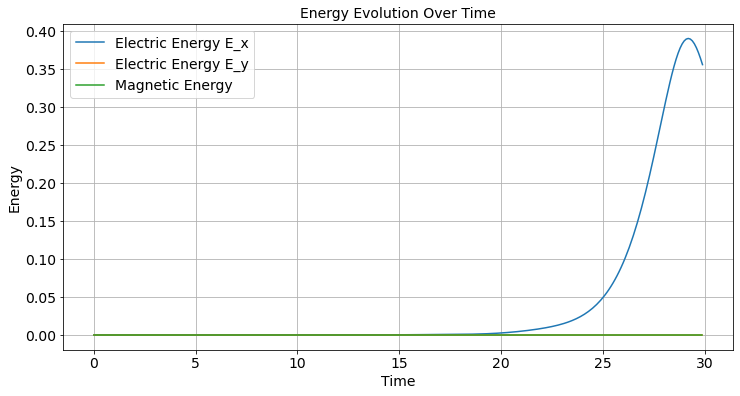}
    \vspace{0.5em}
    {(c) Electromagnetic energies.}
    \label{fig:em_energy_linear}
  \end{minipage}
  \hfill
  \begin{minipage}{0.48\linewidth}
    \centering
    \includegraphics[width=\linewidth]{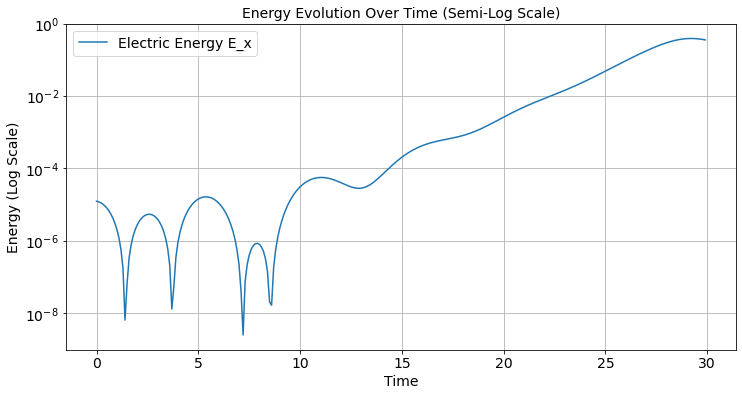}
    \vspace{0.5em}
    {(d) Longitudinal electric energy.}
    \label{fig:em_energy_semilog}
  \end{minipage}
  
  \vspace{1em} 
  
  \begin{minipage}{1.0\linewidth}
    \centering
    \includegraphics[width=\linewidth]{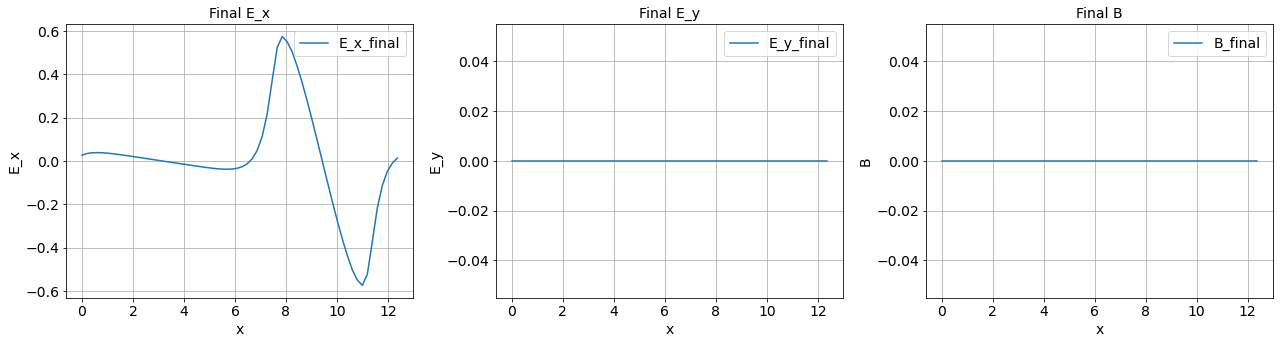}
    \vspace{0.5em}
    {(e) Final electromagnetic fields at \(t=30\).}
    \label{fig:final_fields}
  \end{minipage}
  
  \caption{Results with no control applied: (a) Equilibrium distribution; (b) Distribution at \(t=30\); (c) Electromagnetic energies  (linear scale) from \(t=0\) to \(t=30\); (d) Longitudinal electric energy (semi-log scale) from \(t=0\) to \(t=30\); (e) Final electromagnetic fields at \(t=30\) for $x\in [0, L]$. }
  \label{fig:two_stream_full_presentation_no_control}
\end{figure}

The example we use is the following two-stream instability (also known as the streaming Weibel instability; see, e.g.,~\cite{cheng2014energy,cheng2014discontinuous}). The equilibrium distribution is given by

\begin{equation}
\label{eq:two-stream_equilibrium}
\mu^{\mathrm{two\text{-}stream}}(v_x, v_y)
= \frac{1}{2\pi v_{th}^2}
\left[
\exp\!\left(-\frac{(v_x - \bar{v})^2 + v_y^2}{v_{th}^2}\right)
+
\exp\!\left(-\frac{(v_x + \bar{v})^2 + v_y^2}{v_{th}^2}\right)
\right],
\end{equation}
and we have chosen \(\bar{v}=0.7\) and \(v_{th}=0.2\). We consider a small sine perturbation in \(x\) with amplitude \(\alpha=10^{-3}\), so that the initial distribution is given by
\begin{equation}
     \label{eq:two-stream_instability_perturbation}
     f_{\text{ini}}(x,v_x,v_y) = \mu^{\mathrm{two-stream}}(v_x, v_y) \left[1 + \alpha \sin(kk_0 x)\right].
\end{equation}
We set \(k_0=0.5\) (so that \(L=4\pi\)) and \(k=1\). In Figure \ref{fig:two_stream_full_presentation_no_control}, we plot the  marginal density \(\int_\mathbb{R} f(t,x,v_x,v_y)\,dv_y\) at $t=0$ and $t=30$. At $t=30$ the instability has already saturated. We clearly observe that the longitudinal field $E_x$ grows exponentially in time, while the transverse fields $(E_y,B_z)$ are not significantly modified from their zero initial value.

Since the instability in this case is longitudinal, we cannot use a linear control strategy. We therefore resort to a PDE-constrained optimization approach for designing a control strategy. Specifically, we solve the following PDE-constrained optimization problem

\begin{equation}\label{eqn:optimization}
\begin{aligned}
\min_{\eyexteriorinitial,\bzexteriorinitial}\quad & J\bigl[f(\eyexteriorinitial,\bzexteriorinitial)\bigr], \\[1mm]
\text{s.t.}\quad & \left\{
\begin{aligned}
&\partial_t f + v_x\,\partial_x f + \bigl(E_x + v_y B_z\bigr) \partial_{v_x} f + \bigl(E_y - v_x B_z\bigr) \partial_{v_y} f = 0, \\[1mm]
&\partial_x E_x = \rho {- 1} \\[1mm]
&\partial_t E_y = -\,\partial_x B_z - j_y, \\[1mm]
&\partial_t B_z = -\,\partial_x E_y, \\[1mm]
&f(0,x,v_x,v_y) = f_{\mathrm{ini}}(x,v_x,v_y), \\[1mm]
&{E_y(0,x) = \Eyplasmaini(x)+ \eyexteriorinitial(x), \quad B_z(0,x) = \Bzplasmaini(x)+\bzexteriorinitial(x).}
\end{aligned}
\right.
\end{aligned}
\end{equation}
The goal of the optimization is to determine the initial external fields \((\eyexteriorinitial,\bzexteriorinitial)\) that minimize the instability. There are various ways of defining the objective function \( J \) that quantitatively measures the severity of the instability. In this paper we study $J=J_E$:
\begin{equation}\label{eq:J_definition_electric_x_energy}
 J_E = \int_0^T \left\| E_x(t,\cdot) \right\|^2 _{L^2}\,dt\,.
\end{equation}

We use a truncated variant of \eqref{eq:E_y_B_z_exterior_para_formula} to parametrize {\( (\eyexteriorinitial,\,\bzexteriorinitial) \)}, i.e.
\begin{align*}
  {\eyexteriorinitial(x)} &= \sum_{k\in K} \Bigl[(a_k + d_k) \cos(kk_0 x) + (b_k + c_k) \sin(kk_0 x)\Bigr], \\
  {\bzexteriorinitial(x)} &= \sum_{k\in K} \Bigl[(a_k - d_k) \cos(kk_0 x) + (b_k - c_k) \sin(kk_0 x)\Bigr],
\end{align*}
where \(K\) is a chosen subset of \(\mathbb{Z}^+\) corresponding to the modes we wish to include. The optimization is then performed over the set of parameters \(\{(a_k, b_k, c_k, d_k)\}_{k\in K}\). This helps the optimization, both by reducing the total number of parameters as well as directly encoding the physical constraints on {\( (\eyexteriorinitial,\,\bzexteriorinitial) \)}.

We now conduct the numerical optimization of~\eqref{eqn:optimization}. 
{The forward PDE solver we used here is described in detail in Section \ref{subsec:forward_solver}}. The computational domain is $[0, {L}]\times[-3,3]\times[-3,3]$ and the simulation is run up to $t=30$. The optimization solver is gradient based with a learning-rate $h$ and we update the parameters as follows:
\[
a_k \leftarrow a_k - h\frac{\partial J}{\partial a_k},\quad b_k \leftarrow b_k - h\frac{\partial J}{\partial b_k},\quad c_k \leftarrow c_k - h\frac{\partial J}{\partial c_k},\quad d_k \leftarrow d_k - h\frac{\partial J}{\partial d_k}\,.
\]
Here $h$ can be a fixed value as in the standard gradient descent, or chosen through line-search. To compute the gradient we use the autodifferentiation features of the JAX package \cite{bradbury2018jax, chandrasekhar2021auto,schoenholz2020jax}. Additional details on the optimization solver are provided in Appendix~\ref{appen:sec:optimization_detail}.

\begin{figure}[htbp]
  \centering
  \begin{minipage}{0.6\linewidth}
    \centering
    \includegraphics[width=\linewidth]{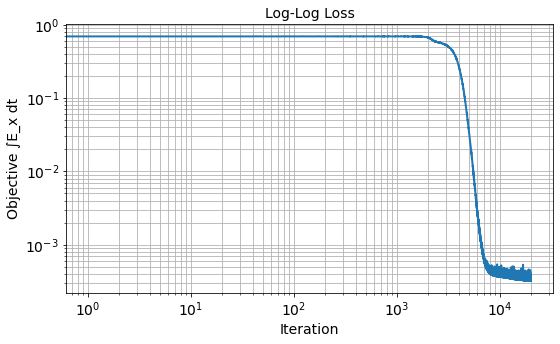}
  \end{minipage}
  
  \caption{Evolution of the objective function {\(J=J_E\)} {given by \eqref{eq:J_definition_electric_x_energy} at $ T = T_{\mathrm{end}}=30$} in log-log scale.  
  }
  \label{fig:two_stream_loss_loglog}
\end{figure}

The to-be-inferred coefficients are \(\{a_k,b_k,c_k,d_k\}_{k\in K}\) with \(K = \{1,2,3,4,5\}\), so there are $20$ parameters in total. We now have multiple modes that can interact nonlinearly. We observe that the optimization algorithm decreases the objective function by more than three orders of magnitude (see Figure \ref{fig:two_stream_loss_loglog}), demonstrating the effectiveness of the approach. The final output for the parameters is given in Table~\ref{tab:best_params}.

\begin{table}[H]
\centering
\caption{Optimized harmonic coefficients $(a_k,b_k,c_k,d_k)$ using the best-so-far parameters up to epoch $10,000$. Objective $J_E=\int_0^{T_{\mathrm{end}}}\|E_x(t,\cdot)\|^2_{L^2}\,dt$ evaluates to $3.823\times 10^{-4}$ at low resolution $(16\times32\times32)$ and $1.879\times 10^{-2}$ at high resolution $(64\times128\times128)$.}
\begin{tabular}{|c|c|c|c|c|}
\toprule
$k$ & $a_k$ & $b_k$ & $c_k$ & $d_k$ \\
\midrule
1 & $-1.670\times10^{-2}$ & $-1.202\times10^{-2}$ & $+1.134\times10^{-2}$ & $-1.729\times10^{-2}$ \\
2 & $-2.089\times10^{-2}$ & $+4.536\times10^{-3}$ & $+5.735\times10^{-3}$ & $+2.174\times10^{-2}$ \\
3 & $+1.306\times10^{-3}$ & $+2.609\times10^{-3}$ & $-1.164\times10^{-3}$ & $-2.565\times10^{-4}$ \\
4 & $-2.175\times10^{-3}$ & $+9.804\times10^{-4}$ & $-1.003\times10^{-3}$ & $+1.205\times10^{-3}$ \\
5 & $-8.161\times10^{-4}$ & $+3.470\times10^{-4}$ & $+9.051\times10^{-5}$ & $+7.656\times10^{-4}$ \\
\bottomrule
\end{tabular} \label{tab:best_params}
\end{table}

\begin{figure}
  \centering
  \begin{minipage}{0.48\linewidth}
    \centering
    \includegraphics[width=\linewidth]{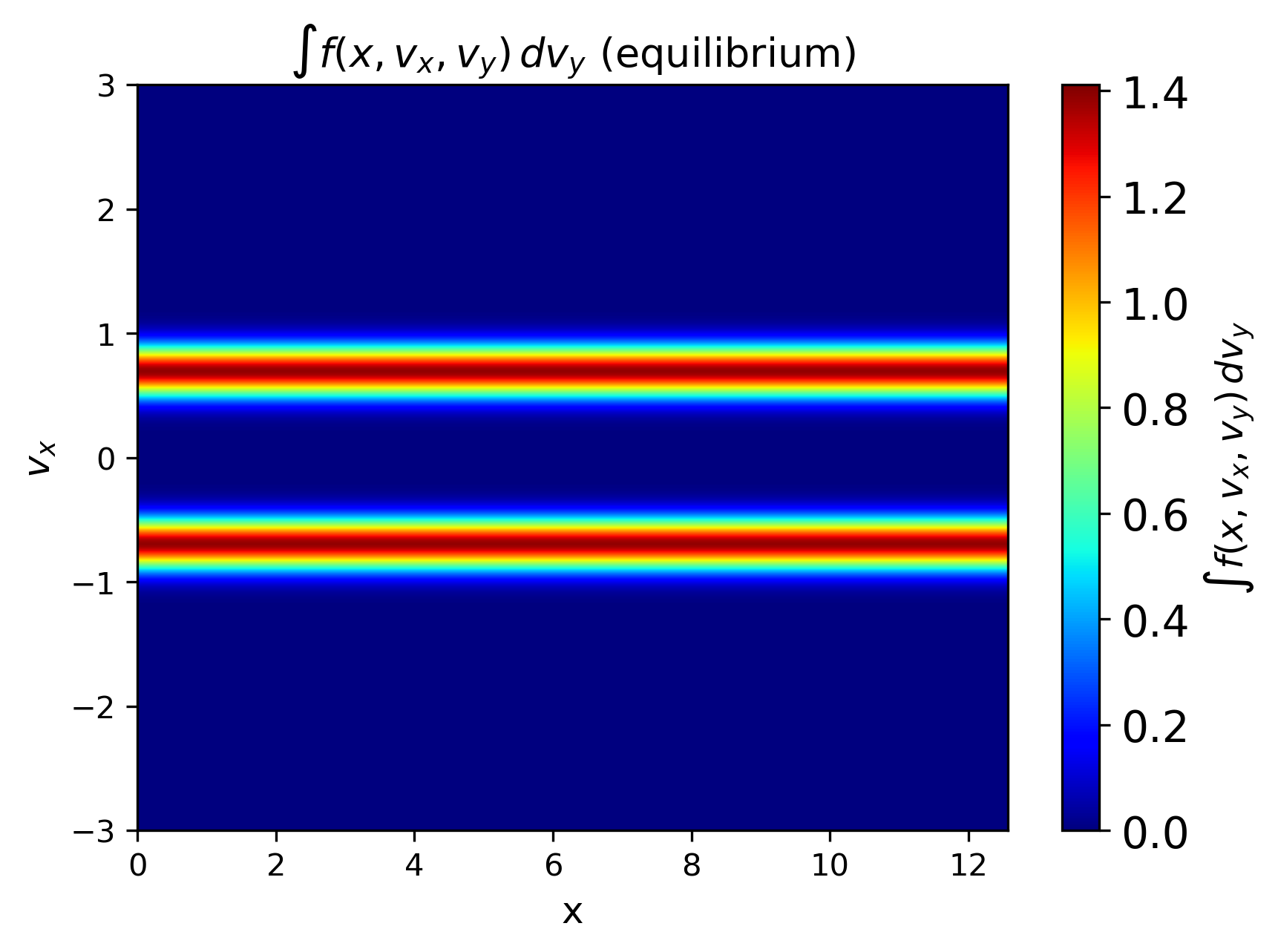}
    \vspace{0.5em}
    {(a) Equilibrium distribution.}
  \end{minipage}
  \hfill
  \begin{minipage}{0.48\linewidth}
    \centering
    \includegraphics[width=\linewidth]{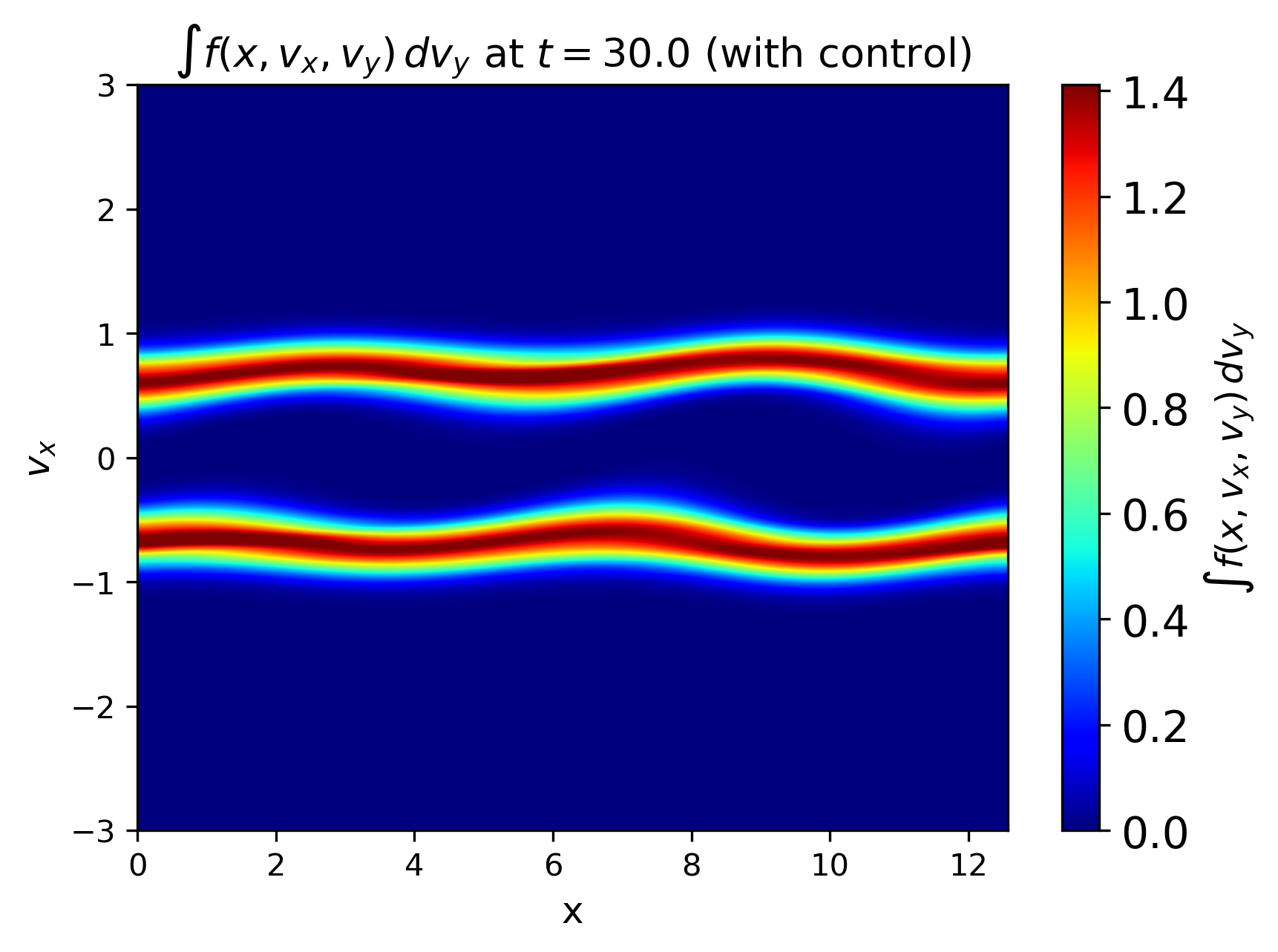}
    \vspace{0.5em}
    {(b) Distribution at \(t=30\).}
    \label{fig:t40_distribution_no_control}
  \end{minipage}
  
  \vspace{1em} 
  
  \begin{minipage}{0.48\linewidth}
    \centering
    \includegraphics[width=\linewidth]{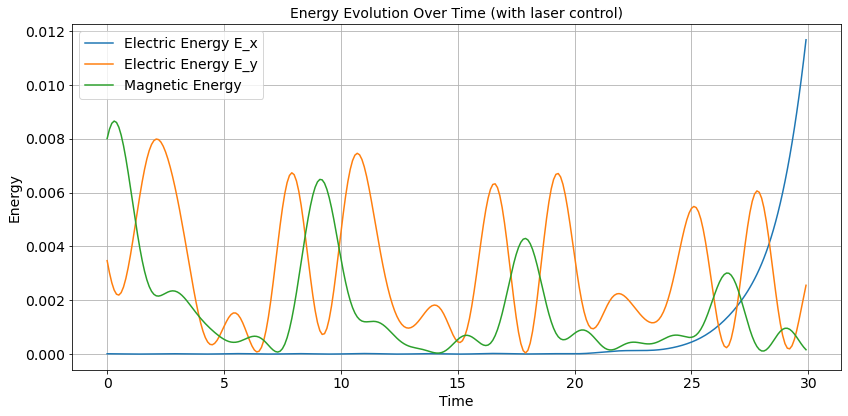}
    \vspace{0.5em}
    {(c) Electromagnetic energies.}
    \label{fig:em_energy_linear}
  \end{minipage}
  \hfill
  \begin{minipage}{0.48\linewidth}
    \centering
    \includegraphics[width=\linewidth]{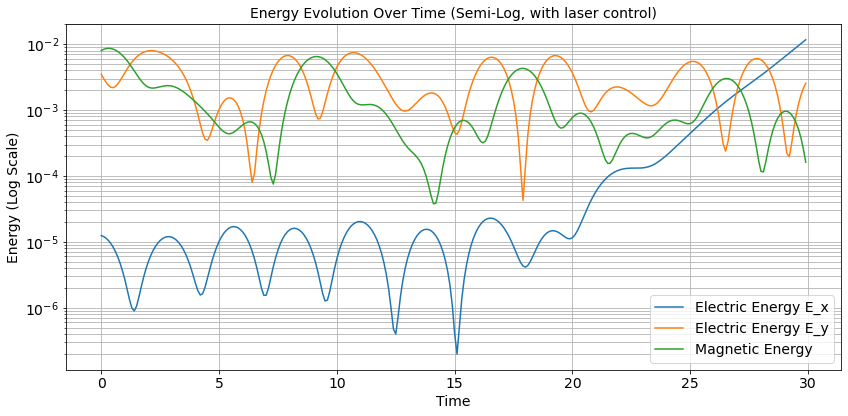}
    \vspace{0.5em}
    {(d) Longitudinal electric energy.}
    \label{fig:em_energy_semilog}
  \end{minipage}
  
  \vspace{1em} 
  
  \begin{minipage}{1.0\linewidth}
    \centering
    \includegraphics[width=\linewidth]{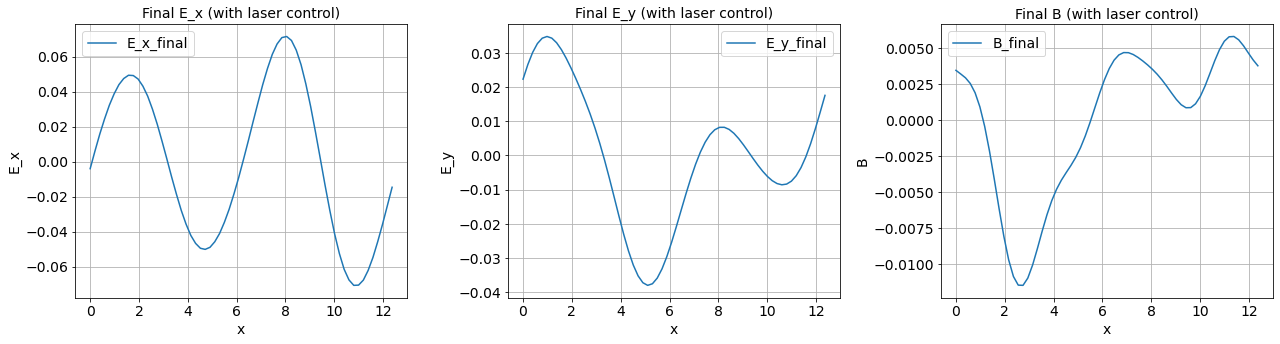}
    \vspace{0.5em}
    {(e) Final electromagnetic fields at \(t=30\) for $x\in [0, L]$.}
    \label{fig:final_fields}
  \end{minipage}
  
  \caption{Results for the optimized parameter set in Table \ref{tab:best_params}.: (a) Equilibrium distribution; (b) Distribution at \(t=30\); (c) Electromagnetic energies  (linear scale) from \(t=0\) to \(t=30\); (d) Longitudinal electric energy (semi-log scale) from \(t=0\) to \(t=30\); (e) Final electromagnetic fields at \(t=30\). }
  \label{fig:two_stream_full_presentation_best_control}
\end{figure}

Since, due to computational constraints, the optimization parameters are found with the forward solver run with a lower resolution, upon getting the optimized parameters, we re-run the Vlasov-Maxwell equations for these specific parameter set with the finer resolution and plot the result in Figure \ref{fig:two_stream_full_presentation_best_control}. In comparison to the uncontrolled scenario (Figure \ref{fig:two_stream_full_presentation_no_control}), the new dynamics shows that the optimized parameters successfully delay the growth of \(E_x\), maintaining its value at a low level at $t=30$. Specifically,  subplot (b) reveals that we preserve the two-stream structure. At \(t=30\), as shown in subplot (e), the maximum value of \(E_x\) is around $0.07$ when control is applied, whereas in the uncontrolled case the maximum of \(E_x\) is approximately $0.58$. Moreover, the field amplitudes for \((E_y,B_z)\) remain small, although not zero as in the uncontrolled case, even though we have not explicitly optimized for this outcome.

\begin{figure}[ht]
    \centering
    \includegraphics[width=0.6\linewidth]{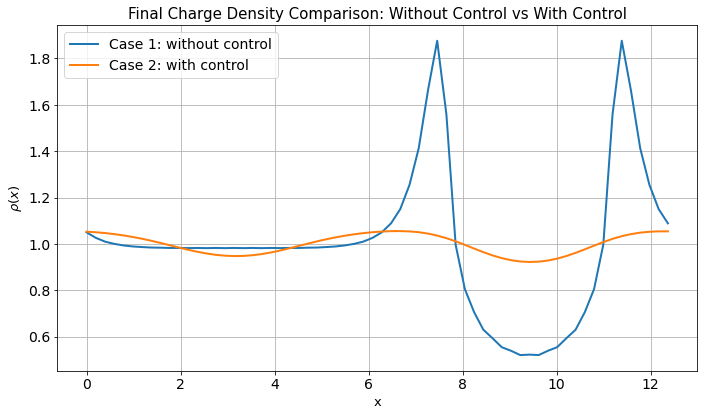}
    \caption{Comparison of density function with and without control.}
    \label{fig:density_comparison}
\end{figure}

To directly compare the controlled and uncontrolled case, we also plot the density function \(\rho(t=30,x)\) in Figure~\ref{fig:density_comparison}. {Under the optimized control, the density \(\rho\) remains close to its equilibrium value \(\rho(x)\equiv 1\) for all \(x\)}, exhibiting substantially smaller deviations than in the uncontrolled one. This marked reduction in density fluctuations confirms that the optimized external fields effectively suppresses the instability over the time interval considered for the optimization. 

\section{Conclusion}

In this paper, we have investigated the control of kinetic plasma instabilities through low-intensity laser fields. First, by extending the classical Penrose criterion to the 1.5D Vlasov–Maxwell system, we derive a generalized stability condition and demonstrate that transverse (Weibel-type) instabilities can be suppressed via appropriately designed laser electromagnetic fields in the linear regime (assuming the perturbation is known). Second, recognizing the limitations of linear theory for longitudinal modes, we have formulated a PDE-constrained optimization approach and applied it to the two-stream instability, showing numerically that carefully tuned laser parameters can effectively delay the onset of the instability by exploiting nonlinear effects. Looking forward, this framework establishes a more realistic pathway for translating theoretical insights and numerical predictions into experimental studies than has been achieved in previous work aimed at mitigating kinetic instabilities. The numerical approach, in particular, enables the optimization of control strategies for broader classes of problems with complex plasma phenomena.

\medskip

\bibliographystyle{abbrv}
\bibliography{reference,shumlak_bibtex}

\appendix

\section{Fourier and Laplace transform}

\subsection{Technical Lemmas}

\begin{lem}\label{lem:xf_Fourier}
    Fourier transform of $xf(x)$ is $i\partial_k\hat{f}(k)$ where $\hat{f}(k)$ is the Fourier transform of $f$ if $f$ has exponential decay at infinity.
\end{lem}
\begin{proof}
    \begin{equation*}
        \widehat{xf(x)}(k)=\realint xf(x)e^{-ikx} \,dx=-\frac{1}{i}\realint\frac{d}{dk}\left(f(x)e^{-ikx}\right)\,dx=i\frac{d}{dk}\left[\realint f(x)\,e^{-ikx}\,dx\right]=i\partial_k\hat{f}(k). 
    \end{equation*}
\end{proof}

\begin{lem}\label{lem:xf'_Fourier}
    Fourier transform of $xf'(x)$ is $-\hat{f}(k)-k\partial_k \hat{f}(k)$. 
\end{lem}
\begin{proof}
    By the previous Lemma, we know that 
    $$\widehat{xf'(x)}(k)=i\partial_k \left(ik\hat{f}(k)\right)= -\hat{f}(k)-k\partial_k \hat{f}(k). 
    $$
\end{proof}
\begin{lem}
    \label{lem:time_derivative_Laplace}
    If $\lim\limits_{t\to\infty} f(t)=0$, then Laplace transform of $\partial_t f$ is $f(0)+sL[f]$ when $s>0$.
\end{lem}
\begin{proof}
    Using integration by parts, we have
    \begin{equation*}
        L[\partial_t f](s)=\int_0^\infty e^{-st}\partial_tf(t)\,dt=e^{-st}f(t)\Big|_0^\infty+s\int_0^\infty f(t)e^{-st}\,dt=-f(0)+sL[f](s).
    \end{equation*}
\end{proof}

\begin{lemma}
    \label{appen:mu_hat_integral}
    Given any mode \(m\), the following properties hold:
    \begin{equation}\label{eq:mu_hat_integral}
        \int_{\mathbb{R}} \widehat{\partial_{v_y} \mu}(m,v_y)\,dv_y = 0,
    \end{equation}
    and 
    \begin{equation}
        \label{eq:mu_hat_v_integral}
        \int_{\mathbb{R}} v_y\widehat{\partial_{v_y} \mu}(m,v_y)\,dv_y = -\int_{\mathbb{R}}\hat{\mu}(m,v_y)\,dv_y,
    \end{equation}
    where \(\widehat{\cdot}\) denotes the Fourier transform with respect to the first velocity coordinate, \(v_x\), and \(\mu\) represents the equilibrium function, assumed to exhibit exponential decay at infinity.
\end{lemma}

\begin{proof}
    To prove \eqref{eq:mu_hat_integral}, we explore the Fourier transform's explicit form concerning the first velocity coordinate:
    \begin{equation*}
        \int_{\mathbb{R}} \widehat{\partial_{v_y} \mu}(m,v_y)\,dv_y = \int_{\mathbb{R}}\int_{\mathbb{R}} \partial_{v_y} \mu(v_x,v_y)e^{-imv_x}\,dv_x\,dv_y = \int_{\mathbb{R}} e^{-imv_x}\left[\mu(v_x,v_y)\Big|_{-\infty}^{\infty}\right]\,dv_x = 0,
    \end{equation*}
    due to the exponential decay of \(\mu\) at infinity.

    Similarly, for \eqref{eq:mu_hat_v_integral}, we find:
    \begin{equation*}
        \int_{\mathbb{R}} v_y\widehat{\partial_{v_y} \mu}(m,v_y)\,dv_y = \int_{\mathbb{R}}\int_{\mathbb{R}} v_y\partial_{v_y} \mu(v_x,v_y)e^{-imv_x}\,dv_x\,dv_y = -\int_{\mathbb{R}}\int_{\mathbb{R}} \mu(v_x,v_y)e^{-imv_x}\,dv_x\,dv_y = -\int_{\mathbb{R}}\hat{\mu}(m,v_y)\,dv_y.
    \end{equation*}
    This completes the proof.
\end{proof}

\subsection{Identities}

\begin{lemma}\label{lem:g_formula}
The function \( g \) satisfies
\begin{equation}\label{eq:g_formula}
\begin{aligned}
\hat{g}(t,k,kt,v_y) - \hat{g}(0,k,kt,v_y)
&+ \int_0^t \hat{\rho}(s,k)(t-s)\,\hat{\mu}\bigl(k(t-s),v_y\bigr)\,ds\\[1mm]
&= -ikv_y\int_0^t \hat{B}_z(s,k)(t-s)\,\hat{\mu}\bigl(k(t-s),v_y\bigr)\,ds\\[1mm]
&\quad -\frac{i}{k}\,\hat{B}_z(t,k)\,\widehat{\partial_{v_y}\mu}(0,v_y)
+\frac{i}{k}\,\hat{B}_z(0,k)\,\widehat{\partial_{v_y}\mu}(kt,v_y).
\end{aligned}
\end{equation}
\end{lemma}

\subsection{Proof of Lemma \ref{lem:g_formula}}\label{appen:Lemma_density}

\begin{proof}

We apply the Fourier transform in \(x\) and \(v_x\) to equation \eqref{eq:g_reformulation_VM}. This yields
\begin{equation}\label{eq:g_relation_Fourier}
\begin{aligned}
    \partial_t \hat{g}(t,k,m,v_y) &+ \iint \Bigl[ E_x\bigl(t,x+v_xt\bigr) + v_y\,B_z\bigl(t,x+v_xt\bigr) \Bigr] \partial_{v_x}\mu(v_x,v_y)\, e^{-ikx}e^{-imv_x}\,dx\,dv_x\\[1mm]
    &\quad + \iint \Bigl[ E_y\bigl(t,x+v_xt\bigr) - v_x\,B_z\bigl(t,x+v_xt\bigr) \Bigr] \partial_{v_y}\mu(v_x,v_y)\, e^{-ikx}e^{-imv_x}\,dx\,dv_x=0,
\end{aligned}
\end{equation}
where \(k\) and \(m\) are the Fourier frequency variables corresponding to \(x\) and \(v_x\), respectively.

To investigate this further, we now calculate each term in \eqref{eq:g_relation_Fourier} separately. For the term involving \(E_x\) and \(\partial_{v_x}\mu\), we have
\begin{align*}
&\iint E_x\bigl(t,x+v_xt\bigr)\,\partial_{v_x}\mu(v_x,v_y)\,e^{-ikx}e^{-imv_x}\,dx\,dv_x\\[1mm]
&\quad = \int \left[\int E_x\bigl(t,x+v_xt\bigr)e^{-ik(x+v_xt)}\,dx\right]\partial_{v_x}\mu(v_x,v_y)e^{-i(m-kt)v_x}\,dv_x\\[1mm]
&\quad = i(m-kt)\,\hat{E}_x(t,k)\,\hat{\mu}(m-kt,v_y).
\end{align*}
For the term involving \(v_yB_z\) and \(\partial_{v_x}\mu\):  similarly, the following calculation holds
\begin{align*}
&\iint v_y\,B_z\bigl(t,x+v_xt\bigr)\,\partial_{v_x}\mu(v_x,v_y)\,e^{-ikx}e^{-imv_x}\,dx\,dv_x\\[1mm]
&\quad = v_y\int \left[\int B_z\bigl(t,x+v_xt\bigr)e^{-ik(x+v_xt)}\,dx\right]\partial_{v_x}\mu(v_x,v_y)e^{-i(m-kt)v_x}\,dv_x\\[1mm]
&\quad = iv_y(m-kt)\,\hat{B}_z(t,k)\,\hat{\mu}(m-kt,v_y).
\end{align*}
For the term involving \(E_y\) and \(\partial_{v_y}\mu\), we obtain
\begin{align*}
&\iint E_y\bigl(t,x+v_xt\bigr)\,\partial_{v_y}\mu(v_x,v_y)\,e^{-ikx}e^{-imv_x}\,dx\,dv_x = \hat{E}_y(t,k)\,\widehat{\partial_{v_y}\mu}(m-kt,v_y).
\end{align*}
For the term involving \(v_xB_z\) and \(\partial_{v_y}\mu\),  utilizing Lemma \ref{lem:xf_Fourier}, we derive
\begin{align*}
&\iint v_x\,B_z\bigl(t,x+v_xt\bigr)\,\partial_{v_y}\mu(v_x,v_y)\,e^{-ikx}e^{-imv_x}\,dx\,dv_x\\[1mm]
&\quad = \int \left[\int B_z\bigl(t,x+v_xt\bigr)e^{-ik(x+v_xt)}\,dx\right] v_x\,\partial_{v_y}\mu(v_x,v_y)e^{-i(m-kt)v_x}\,dv_x\\[1mm]
&\quad = \hat{B}_z(t,k)\,\widehat{v_x\,\partial_{v_y}\mu}(m-kt,v_y)\\[1mm]
&\quad = i\hat{B}_z(t,k)\,\partial_m\widehat{\partial_{v_y}\mu}(m-kt,v_y),
\end{align*}
where \(\partial_m\) denotes differentiation with respect to the Fourier variable \(m\).

On the other hand, the Fourier transforms of the Poisson equation \eqref{eq:Poisson_no_pert} and the magnetic field equation \eqref{eq:magnetic_field_no_pert} yield
\begin{equation}\label{eq:Poisson_Fourier}
ik\,\hat{E}_x(t,k) = \hat{\rho}(t,k),
\end{equation}
\begin{equation}\label{eq:Bz_Fourier}
\partial_t \hat{B}_z(t,k) = -ik\,\hat{E}_y(t,k).
\end{equation}
Furthermore, applying the Fourier transform to equations \eqref{eq:tranverse_electric_field_no_pert} and \eqref{eq:current_traverse} gives
\begin{equation}\label{eq:traverse_electric_field_Fourier}
\begin{aligned}
\partial_t \hat{E}_y(t,k) &= -ik\,\hat{B}_z(t,k) - \iiint v_y\,f(t,x,v_x,v_y)e^{-ikx}\,dv_x\,dv_y\,dx\\[1mm]
&=-ik\,\hat{B}_z(t,k)-\int_{\mathbb{R}} v_y\,\hat{g}(t,k,kt,v_y)\,dv_y.
\end{aligned}
\end{equation}

Starting from equation \eqref{eq:g_relation_Fourier} and incorporating the term-by-term calculations as well as the Fourier-transformed field equations \eqref{eq:Poisson_Fourier}-\eqref{eq:traverse_electric_field_Fourier}, we express \(\hat{g}\) in terms of the density \(\hat{\rho}\) and the magnetic field \(\hat{B}_z\). Given \eqref{eq:rho_Fourier_detailed}, the density function correlates with $\hat{g}$ through its integration over $v_y$ with the mode for longitudinal velocity selected as $kt$. Integrating \eqref{eq:g_relation_Fourier}  from $0$ to $t$ and set $m=kt$, we arrive at
\begin{equation}\label{eq:g_inter2}
    \begin{aligned}
        &\quad\,\,\hat{g}(t,k,kt,v_y) - \hat{g}(0,k,kt,v_y)\\
        &= -\int_0^t \hat{\rho}(s,k)(t-s)\hat{\mu}\left(k(t-s),v_y\right)\,ds-ik\int_0^t v_y(t-s)\hat{B}_z(s,k)\hat{\mu}\left(k(t-s),v_y\right)\,ds\\
        &\quad\,\,-\frac{i}{k}\int_0^t \partial_s\hat{B}_z(s,k)\widehat{\pvtwo\mu}\left(k(t-s),v_y\right)\,ds+ i\int_0^t \hat{B}_z(s,k)\partial_m\widehat{\pvtwo\mu}\left(k(t-s),v_y\right)\,ds
    \end{aligned}
\end{equation}
The third term demands further scrutiny. An integration-by-parts approach yields
\begin{align*}
    &\quad\,\,\frac{i}{k}\int_0^t \partial_s\hat{B}_z(s,k)\widehat{\pvtwo\mu}\left(k(t-s),v_y\right)\,ds\\
    & = \frac{i}{k}\int_0^t \widehat{\pvtwo\mu}\left(k(t-s),v_y\right)\,d\,\hat{B}_z(s,k)\\
    & = \frac{i}{k}\,\hat{B}_z(s,k)\,\widehat{\pvtwo\mu}\left(k(t-s),v_y\right)\Big|_0^t-\frac{i}{k}\int_0^t \hat{B}_z(s,k)\,d\,\left[\widehat{\pvtwo\mu}\left(k(t-s),v_y\right)\right]  \\
    &=\frac{i}{k}\hat{B}_z(t,k)\,\widehat{\pvtwo\mu}(0,v_y)-
   \frac{i}{k}\hat{B}_z(0,k)\,\widehat{\pvtwo\mu}(kt,v_y)+i\int_0^t\hat{B}_z(s,k)\,\partial_m\widehat{\pvtwo\mu}(k(t-s),v_y)\,ds.
\end{align*}
This calculation nicely cancels the last term in \eqref{eq:g_inter2} and it also introduces the boundary terms of $\hat{B}$. Incorporating this into \eqref{eq:g_inter2} completes the proof of the desired result \eqref{eq:g_formula}.
\end{proof}

\subsection{Laplace Transform of Density Function}\label{appen:lemma_density_Laplace}

\begin{lemma}\label{appen:lemma:density_Laplace}
    The Laplace transform of the density function, $\hat{\rho}$, adheres to the relation:
    \begin{equation}\label{appen:rho_final}
        \begin{aligned}
            L[\hat{\rho}(\cdot, k)](s) \left[1 + L[\Geq(\cdot, k)](s) \right] +ik L[\hat{B}_z(\cdot, k)](s) \,L[\Geqv(\cdot, k)](s)  =  L[\Ginit(\cdot, k)](s).
        \end{aligned}
    \end{equation}
   
\end{lemma}

\begin{proof}
By integrating \eqref{eq:g_formula} over $v_y$ and employing \eqref{eq:rho_Fourier_detailed}, we derive the following relationship for $\hat{\rho}(t,k)$:
\begin{equation*}
    \begin{aligned}
        \hat{\rho}(t,k) &= \int_{\mathbb{R}} \hat{g}(t,k,kt,v_y)\,dv_y\\
        &= \int_{\mathbb{R}} \hat{g}(0,k,kt,v_y)\,dv_y - \int_0^t \hat{\rho}(s,k)\left[(t-s)\realint\hat{\mu}\left(k(t-s),v_y\right)\,dv_y\right]\,ds\\
         &\quad -ik\int_0^t \hat{B}_z(s,k)(t-s)\left[\int_{\mathbb{R}}v_y\hat{\mu}\left(k(t-s),v_y\right)\,dv_y\right]\,ds\\
         &\quad-\frac{i}{k}\hat{B}_z(t,k)\,\realint\widehat{\pvtwo\mu}(0,v_y)\,dv_y+
   \frac{i}{k}\hat{B}_z(0,k)\,\realint\widehat{\pvtwo\mu}(kt,v_y)\,dv_y
    \end{aligned}
\end{equation*}
Using Lemma \ref{appen:mu_hat_integral}, the last two terms involving \(\widehat{\partial_{v_y}\mu}\) are both zero. Recognizing a convolution structure within the other terms, we apply the Laplace transform, which translates convolutions into products \cite{abell2018introductory}. This approach, when applied to each term for distinct values of \(k\), yields:
\begin{equation*}
    \begin{aligned}
        L[\hat{\rho}(\cdot, k)](s) &= L[\Ginit(\cdot, k)](s) - L[\hat{\rho}(\cdot, k)](s) \,L[\Geq(\cdot, k)](s)- ikL[\hat{B}_z(\cdot, k)](s) L[\Geqv(\cdot, k)](s).
    \end{aligned}
\end{equation*}
Reorganizing these terms leads to the formulation of the desired result.
\end{proof}

\subsection{Laplace Transform Involving $B_z$}\label{append:lemma_current_relation}

\begin{lemma}\label{appen:lem:current_relation}
    The relationship below is established:
    \begin{equation}\label{eq:current_final}
    \begin{aligned}
   &\quad\,\,ikL[\hat{\rho}(\cdot,k)](s)\,L[\Geqv(\cdot,k)](s)+L[\hat{B}_z(\cdot,k)](s)\left[k^2+{s^2}+1-k^2L\left[\Geqvtwo(\cdot,k)\right](s)\right] \\
       &=-ik\hat{E}_y(0,k)+\hat{B}_z(0,k)\left[s+L\left[\Geqzero(\cdot,k)\right](s) 
 \right] +ik L\left[ \Ginitv(\cdot,k) \right](s),
   \end{aligned}
   \end{equation}
\end{lemma}

\begin{proof}
Incorporating \(\hat{g}(t,k,kt,v_y)\) from \eqref{eq:g_formula} into \eqref{eq:traverse_electric_field_Fourier}, we deduce
\begin{align*}
    \partial_t \hat{E}_y(t,k) &= -ik\hat{B}_z(t,k) - \int_{\mathbb{R}} v_y\hat{g}(0,k,kt,v_y)\,dv_y\\
    &\quad +\int_0^t \hat{\rho}(s,k)\left[\int_{\mathbb{R}}(t-s) v_y\hat{\mu}\left(k(t-s),v_y\right)\,dv_y\right]\,ds\\
    &\quad +ik\int_0^t \hat{B}_z(s,k)\left[\int_{\mathbb{R}} (t-s)v_y^2\hat{\mu}\left(k(t-s),v_y\right)\,dv_y\right]\,ds\\
    &\quad +\frac{i}{k}\hat{B}_z(t,k)\int_{\mathbb{R}} \vtwo\widehat{\pvtwo\mu}(0,v_y)\,dv_y-\frac{i}{k}\hat{B}_z(0,k)\int_{\mathbb{R}} \vtwo\widehat{\pvtwo\mu}(kt,v_y)\,dv_y.
\end{align*}
By Lemma \ref{appen:mu_hat_integral}, we have
\begin{equation*}
    \realint\vtwo\widehat{\pvtwo\mu}(0,v_y)\,dv_y=-\realint\hat{\mu}(0,\vtwo)\,d\vtwo,\quad\quad\quad \realint\vtwo\widehat{\pvtwo\mu}(kt,v_y)\,dv_y=-\realint\hat{\mu}(kt,\vtwo)\,d\vtwo.
\end{equation*}
It is worth noticing that
\begin{equation*}
\realint\hat{\mu}(0,\vtwo)\,d\vtwo = \realint\realint\mu(\vone,\vtwo)\,d\vone\,d\vtwo=1
\end{equation*}
by definition of equilibrium function $\mu$. 

Substituting these values into the equation above and applying the Laplace transform and utilizing Lemma \ref{lem:time_derivative_Laplace}, for \(s>0\), yields
\begin{align*}
   s L[\hat{E}_y(\cdot,k)](s)-\hat{E}_y(0,k) &= -ikL[\hat{B}_z(\cdot,k)](s) - L[\Ginitv(\cdot,k)](s) + L[\hat{\rho}(\cdot,k)](s)L[\Geqv(\cdot,k)](s)\\
   &\quad + ik L[\hat{B}_z(\cdot,k)](s)\,L[\Geqvtwo(\cdot,k)](s) -\frac{i}{k}L[\hat{B}_z(\cdot,k)]+\frac{i}{k}\hat{B}_z(0,k)\,L[\Geqzero(\cdot,k)](s).
\end{align*}
Implementing the Laplace transform on \eqref{eq:magnetic_field_no_pert} and revisiting Lemma \ref{lem:time_derivative_Laplace}, we find
\begin{equation*}
    sL[\hat{B}_z(\cdot,k)](s)-\hat{B}_z(0,k) = -ikL[\hat{E}_y(\cdot,k)](s).
\end{equation*}

Substituting \(L[\hat{E}_y(\cdot,k)]\) with \(\left[\frac{is}{k}L[\hat{B}_z(\cdot,k)]-\frac{i}{k}\hat{B}_z(0,k)\right]\) in the preceding equation, we derive
\begin{align*}
    &\quad\,\,\frac{is^2}{k}L[\hat{B}_z(\cdot,k)](s) - \frac{is}{k}\hat{B}_z(0,k) - \hat{E}_y(0,k) \\&= -ikL[\hat{B}_z(\cdot,k)](s) - L[\Ginitv(\cdot,k)](s) + L[\hat{\rho}(\cdot,k)](s)\,L[\Geqv(\cdot,k)](s)\\
   &\quad + ik L[\hat{B}_z(\cdot,k)](s)\,L[\Geqvtwo(\cdot,k)](s) -\frac{i}{k}L[\hat{B}_z(\cdot,k)]+\frac{i}{k}\hat{B}_z(0,k)\,L[\Geqzero(\cdot,k)](s).
\end{align*}

Rearranging these terms, we establish the relationship presented in \eqref{eq:current_final}.
\end{proof}

\section{Vanishing Lemma for Laplace Transforms}\label{appen:sec:vanishing_laplace_transform}

\begin{lemma}\label{appen:lem:vanishing_laplace_transform}
Let \( A: [0, \infty) \to \mathbb{R} \) and \( B: [0, \infty) \to \mathbb{R} \) be locally integrable functions. Assume:
\begin{enumerate}
    \item The Laplace transforms \( {L}[A](s) \) and \( {L}[B](s) \) exist for all \( s \in \mathbb{C} \) with \( \Re(s) > 0 \).
    \item \( {L}[A](s) \cdot {L}[B](s) = 0 \) for all \( s \) with \( \Re(s) > 0 \).
    \item \( {L}[A](s) \) has at most finitely many zeros in \( \{s \in \mathbb{C} \mid \Re(s) > 0 \} \).
\end{enumerate}
Then \( B(t) = 0 \) for almost every \( t \geq 0 \).
\end{lemma}

\begin{proof}
Since \( {L}[A](s) \) and \( {L}[B](s) \) are defined for \( \Re(s) > 0 \), they are analytic in the right half-plane \( \Omega = \{s \in \mathbb{C} \mid \Re(s) > 0 \} \), which follows from standard Laplace transform theory \cite[Theorem 6.1]{doetsch2012introduction}.

From the given condition,
\[
{L}[A](s) \cdot {L}[B](s) = 0, \quad \forall s \in \Omega.
\]
Let \( Z = \{s_1, \dots, s_n\} \subset \Omega \) be the finite set of zeros of \( {L}[A] \). For \( s \in \Omega \setminus Z \), we have \( {L}[A](s) \neq 0 \), implying
\[
{L}[B](s) = 0, \quad \forall s \in \Omega \setminus Z.
\]
Since \( \Omega \setminus Z \) is an open connected subset of \( \mathbb{C} \) and \( {L}[B](s) \) is analytic in \( \Omega \), it follows by the identity theorem for holomorphic functions \cite[Theorem 8.1.3]{hille2002analytic} that
\[
{L}\{B\}(s) = 0, \quad \forall s \in \Omega.
\]
Finally, by the uniqueness of the inverse Laplace transform \cite[Theorem 5.1]{doetsch2012introduction}, we conclude that \( B(t) = 0 \) for almost every \( t \geq 0 \).
\end{proof}

\section{Discussion of Optimization}\label{appen:sec:optimization_detail}

In this part, we describe the optimization procedure used to solve problem~\eqref{eqn:optimization}. It is a modified version of Armijo method~\cite{armijo1966minimization}, with added Brownian motion. We parameterize the initial electromagnetic fields with \(K\) harmonics, introducing \(4K\) free parameters
\[
p = (a_1,b_1,c_1,d_1,\dots,a_K,b_K,c_K,d_K) \in \mathbb{R}^{4K},
\]
which define the initial values of \((\eyexteriorinitial,\bzexteriorinitial)\). In our implementation, we set \(K = 5\) to balance computational efficiency with control effectiveness.  The objective function is denoted by \(J(p)\).

The update is
\[
p^{(m+1)} \;=\; p^{(m)} \;-\;\alpha_m\,\nabla J\bigl(p^{(m)}\bigr)\;+\;\xi^{(m)},
\qquad
\xi^{(m)}\sim \mathcal{N}(0,\sigma^2 I),
\]
where \(\xi^{(m)}\) is an added small Gaussian perturbation (with \(\sigma=10^{-7}\), or \(\sigma=0\) to disable noise). The descent direction is \(d = -\nabla J\bigl(p^{(m)}\bigr)\), and the step size \(\alpha_m\) is determined by the following Armijo backtracking procedure.

The whole procedure is:
\begin{enumerate}
  \item Initialize \(\alpha \leftarrow \alpha_0 = 10^{-6}\).
  \item Compute the directional derivative \(s = \nabla J\bigl(p^{(m)}\bigr)^\top d\) with \(d = -\nabla J\bigl(p^{(m)}\bigr)\). Here $d$ is the steepest descent direction.
  \item While
  \[
    J\bigl(p^{(m)} + \alpha\,d\bigr)
    \;>\;
    J\bigl(p^{(m)}\bigr)
    \;+\;
    c\,\alpha\,s
  \]
  (with a maximum of 20 halvings), halve \(\alpha \leftarrow \tfrac12\,\alpha\).
  \item Set \(\alpha_m = \alpha\).
  \item Draw noise \(\xi^{(m)}\sim\mathcal{N}(0,\sigma^2 I)\).
  \item Update
  \[
    p^{(m+1)} \;=\; p^{(m)} \;+\;\alpha_m\,d \;+\;\xi^{(m)}.
  \]
\end{enumerate}
In the implementation we set \(c=10^{-4}\) for the constant, and terminate after \(2\times10^4\) iterations or once \(J(p)<10^{-6}\).

The implementation relies on JAX primitives (\texttt{jit}, \texttt{grad}, \texttt{lax.scan}) for both PDE simulation and optimization. During optimization, the forward solver uses a coarse grid of \(16\times32\times32\) points to reduce memory usage. All other simulations use a finer \(64\times256\times256\) grid, confirming that the optimized parameters remain robust and effective across different mesh resolutions.

\end{document}